\documentclass[11pt]{article}
\pdfoutput=1 
\usepackage{jheppub} 
\usepackage{color} 
\usepackage[normalem]{ulem} 
\usepackage{slashed} 
\usepackage{mathrsfs} 
\usepackage{xparse} 
\usepackage{subcaption}
\usepackage{MnSymbol}
\usepackage{amsmath}
\usepackage{mathtools}
\usepackage{float}
\allowdisplaybreaks
\usepackage{relsize}
\usepackage{upgreek}
\usepackage{verbatim}
\usepackage[export]{adjustbox}

\graphicspath{{./}{./figs/}}

\allowdisplaybreaks

\NewDocumentCommand{\mysumint}{e{^_}}{%
  \sumint
  \IfValueT{#1}{^{#1}}%
  \IfValueT{#2}{_{\,#2}}%
}

\newcommand{\MSbar}{$\overline{\text{MS}}$}
\newcommand{\Tc}{T_\text{c}}

\newcommand{\rr}{\tilde{m}_{3}^2}
\newcommand{\rstar}{\tilde{m}_{3,*}^2}
\newcommand{\drstar}{\rr - \rstar}

\title{Higher orders for cosmological phase transitions:\\ A global study in a Yukawa model}
\date{\today}

\author[a]{Oliver Gould}
\author[b]{and Cheng Xie}

\affiliation[a]{School of Physics and Astronomy, University of Nottingham,\\Nottingham NG7 2RD, United Kingdom}
\affiliation[b]{Dept. of Physics and Astronomy, University College London,\\London WC1E 6BT, United Kingdom}

\emailAdd{oliver.gould@nottingham.ac.uk}
\emailAdd{cheng.xie.19@ucl.ac.uk}

\abstract{
We perform a state-of-the-art global study of the cosmological thermal histories of a simple Yukawa model, and find higher perturbative orders to be important for determining both the presence and strength of strong first-order phase transitions.
Using high-temperature effective field theory, we calculate the free energy density of the model up to $\mathcal{O}(y^5T^4)$, where $y$ is the Yukawa coupling and $T$ is the temperature.
The locations of phase transitions are found using the results of lattice Monte-Carlo simulations, and the strength of first-order transitions are evaluated within perturbation theory, to 3-loop order.
This is the first global study of any model at this order.
Compared to a vanilla 1-loop analysis, accurate to $\mathcal{O}(y^2 T^4)$, reaching such accuracy enables on average a five-fold reduction in the relative error in the predicted critical temperature $T_\text{c}$, and an additional $\sim50\%$ strong first-order transitions with latent heat $L/T_\text{c}^4 > 0.1$ to be identified in our scan.
}

\begin{document}

\maketitle

\section{Introduction}

Extensions of the Standard Model (SM) of particle physics can lead to additional phase transitions in the early universe, with consequences for open problems such as baryogensis and dark matter. This in turn can produce a gravitational wave background detectable at current and future detectors such as LISA \cite{LISA:2017pwj}, Taiji \cite{Ruan:2018tsw}, TianQin \cite{TianQin:2015yph}, DECIGO \cite{Kawamura:2011zz}, NANOGrav \cite{NANOGrav:2023hvm} and the International Pulsar Timing Array \cite{Hobbs:2009yy}. For a sizeable gravitational wave background to be produced requires a first-order phase transition, unlike the smooth crossovers corresponding to electroweak symmetry breaking \cite{DOnofrio:2015gop} and colour confinement \cite{Aoki:2006we} in the SM.

We study perhaps the simplest model giving rise to a first-order phase transition, by coupling a real scalar field ($\phi$) to a Dirac fermion ($\psi$) via a Yukawa interaction,
\begin{equation} \label{eq:lagrangian}
\mathscr{L} = \mathscr{L}_\text{Scalar} + \mathscr{L}_\text{Dirac} + \mathscr{L}_\text{Yukawa} \ ,
\end{equation}

\noindent where, in the mostly minus metric and using the Feynman slash notation, $\slashed{\partial} = \gamma^{\mu} \partial_{\mu}$, 
\begin{align}
    \mathscr{L}_\text{Scalar} &= \frac{1}{2}\partial_\mu \phi \partial^\mu \phi - \sigma \phi - \frac{1}{2}m^2\phi^2 - \frac{1}{3!}g \phi^3 - \frac{1}{4!}\lambda \phi^4 ,
    \\
    \mathscr{L}_\text{Dirac} &= \bar{\psi}(i\slashed{\partial} - m_\psi) \psi , \\
    \mathscr{L}_\text{Yukawa} &=  - y \phi \bar{\psi} \psi .
\end{align}

In principle, this Yukawa theory may be coupled to the SM through the Higgs portal. The resulting model has been widely studied as providing a simple dark matter candidate \cite{Kim:2008pp, Baek:2011aa, Baek:2012uj, Esch:2013rta, Buchmueller:2013dya, Klasen:2013ypa, Esch:2014jpa, Freitas:2015hsa, Baek:2015lna, Dupuis:2016fda, Albert:2016osu, Bell:2016ekl}, and is sometimes known as \textit{singlet fermionic dark matter}. Phase transitions in this model have been studied in refs.~\cite{Fairbairn:2013uta, Li:2014wia, Beniwal:2018hyi}. However, in what follows, we will neglect Higgs portal couplings, and focus on the minimal model of $\phi$ and $\psi$, as our primary interest is the reliability and convergence of perturbative approaches to first-order phase transitions. Our model can still have more direct physical relevance in the context of feebly interacting massive particles \cite{Bernal:2017kxu, Yaguna:2023kyu}.

At the high temperatures necessary for a phase transition, with $\pi T$ large compared to $m$ and $m_\psi$, the usual loop expansion breaks down due to the hierarchy of scales introduced between the temperature and relevant masses. Whereas the mass of the zero Matsubara mode has thermal contributions only at the \textit{soft} scale $\mathcal{O}(yT)$, the nonzero Matsubara modes receive thermal contributions to their masses at the \textit{hard} scale $\mathcal{O}(\pi T)$. This scale hierarchy causes some Feynman diagrams to be larger than suggested by loop counting. Effective field theory, applied in the framework of \textit{high-temperature dimensional reduction} \cite{Farakos:1994kx, Farakos:1994xh, Braaten:1995cm, Kajantie:1995dw, Braaten:1995jr}, offers a systematic means to account for this through resummations, and the only framework which has been pushed to higher perturbative orders. For an introduction, see ref.~\cite{Braaten:1995cm}.

Perturbative calculations of the thermodynamics of cosmological phase transitions have been observed to suffer from huge theoretical uncertainties in a variety of models, often amounting to a factor of between a hundred and a million for the predicted gravitational wave peak amplitude. Numerically, often the largest theoretical uncertainty appears as a strong dependence of physical observables on the renormalisation scale \cite{Croon:2020cgk, Papaefstathiou:2020iag, Gould:2021oba, Athron:2022jyi, Curtin:2022ovx, Kierkla:2022odc}. Relatedly, computations using different approximations have been found to yield drastically different predictions \cite{Carena:2019une, Ellis:2022lft}. However, this uncertainty is significantly reduced with higher order dimensional reduction, demonstrating the importance of such calculations.

High-temperature dimensional reduction has been applied at 2-loop order to a wide range of models, including the Standard Model \cite{Kajantie:1995dw} and a number of its extensions (see for example refs.~\cite{Gorda:2018hvi, Biondini:2022ggt}), and at 3-loop and partial 4-loop order to QCD \cite{Braaten:1995jr, Kajantie:2002wa} and the Standard Model in the electroweak symmetric phase \cite{Gynther:2005dj}. The generic rules of dimensional reduction were put down in ref.~\cite{Kajantie:1995dw}, and the first software package to carry this out, DRalgo, was published recently \cite{Ekstedt:2022bff}. The dimensional reduction of the Yukawa model considered here has not previously been carried out. Our results in fact provided a correction for DRalgo, to be included in an upcoming release.

Motivated by making reliable predictions for gravitational wave experiments, recently there has been growing interest in quantifying and reducing theoretical uncertainties for cosmological phase transitions \cite{Cutting:2019zws, Giese:2020znk, Croon:2020cgk, Jinno:2020eqg, Ajmi:2022nmq, Athron:2022jyi, Sagunski:2023ynd, Athron:2023rfq}.
A number of recent studies have tested the validity and accuracy of perturbation theory, utilising dimensional reduction to push to higher orders \cite{Kainulainen:2019kyp, Croon:2020cgk, Niemi:2020hto, Niemi:2021qvp, Gould:2021oba, Tenkanen:2022tly, Gould:2023ovu}. While these studies showed that lower-order calculations can suffer from large theoretical uncertainties, this was based on studying a small number of benchmark parameter points. One is led to wonder how generic these conclusions are. For example, in a broad scan of the parameter space of a model, do the global features of the results change between successive perturbative orders; in short, \textit{does the blob move?}%
\footnote{We thank Graham White for this particularly succinct phrasing of the question.}
In this work, we aim to address this and related questions for our Yukawa model.

\section{Dimensional reduction}

Equilibrium thermodynamics of quantum field theories can be studied in the imaginary time formalism~\cite{Kapusta:2006pm}. The fields of the partition function are defined in three infinite spatial directions and one compact direction with length $1/T$. In the compact direction, bosons, including the scalar $\phi$ in our model, satisfy periodic boundary conditions and can be expanded as
\begin{align}
\phi(\tau, \mathbf{x}) = \mathlarger{\mathlarger{\sum}}_{n = - \infty}^{\infty} \varphi_n(\mathbf{x}) e^{i 2 n \pi T \tau } \ ,
\end{align}
where $\tau$ parameterise the compact direction and $\mathbf{x}$ the spatial directions. On the other hand, fermions such as $\psi$ satisfy anti-periodic boundary conditions and can be expanded as
\begin{align}
\psi(\tau, \mathbf{x}) = \mathlarger{\mathlarger{\sum}}_{n = - \infty}^{\infty} \uppsi_n(\mathbf{x}) e^{i (2 n+1) \pi T \tau } \ .
\end{align}
These Fourier modes are referred to as Matsubara modes \cite{Matsubara:1955ws}. The free part of the Euclidean equilibrium action can be written as
\begin{align}
S_0 = \frac{1}{T} \int d^3 \mathbf{x} \mathlarger{\mathlarger{\sum}}_{n = - \infty}^{\infty}
\bigg[&
\frac{1}{2} (\partial_i \varphi_n)^2 + \frac{1}{2} (2 \pi T n)^2 \varphi_n^2 + \frac{1}{2} m^2 \varphi_n^2
\nonumber \\
&+\bar{\uppsi}_n\left(\gamma_i \partial_i + i 2\pi T (n+1)\gamma_0 + m_\psi \right) \uppsi_n
\bigg] \ , \label{eq:S0}
\end{align}
where $\gamma_i$ and $\gamma_0$ are Euclidean gamma matrices \cite{Laine:2016hma}.

Thus one can view the equilibrium thermodynamics of the four dimensional (4d) theory as the vacuum dynamics of infinitely many 3d fields \cite{Andersen:2004fp}, the fields $\varphi_n$ with effective squared masses $m_n^2 = m^2 + (2 \pi T n)^2$\, and the fields $\uppsi_n$ with effective squared masses $m_\psi^2 + (2\pi T (n+1))^2$.

At temperature $T$ the effective coupling constant of the $n=0$ light bosonic mode increases as $\propto T/m_0$, reflecting its high occupancy $n_\text{B} \sim T / m_0$. The naive perturbative expansion for this mode therefore breaks down at high temperatures. A resummed perturbative expansion is nevertheless possible, though its convergence is slower than at zero temperature.

At high temperatures some form of resummation is necessary, yet the specifics depend on what assumptions are made regarding the relative magnitudes of different parameters \cite{Lofgren:2023sep}.
In this paper, the following power counting prescriptions were adopted for the parameters of the Yukawa theory, fixing the relative sizes of the couplings such that: (i) the loop expansion parameters at zero temperature are of similar size, and (ii) the thermal contributions to the effective potential are of similar size to the tree-level terms, as expected for a phase transition (additional details given in appendix \ref{appendix: power counting}): 
\begin{equation} \label{eq: power counting prescription}
    \frac{\sigma}{T^3}
    \sim \frac{m^2}{T^2}
    \sim \frac{g}{T}
    \sim \lambda
    \sim y^2
    \sim \frac{m_\psi^2}{T^2}  \ .
\end{equation}

Using the hierarchy of scales between masses and the temperature, we can construct an effective field theory (EFT) for only the light modes. This construction is referred to as \emph{dimensional reduction} due to the EFT being defined in three dimensions, and is the most general 3d theory obeying the same internal and spatial symmetries as the 4d theory, as well as having the same number of light bosonic degrees of freedom:
\begin{align} \label{eq: 3d Lagrangian}
    \mathscr{L}_\text{3} &= \frac{1}{2}(\partial_i \phi_3)^2 + \sigma_3 \phi_3 + \frac{1}{2}m_3^2\phi_3^2 + \frac{1}{3!}g_3 \phi_3^3 + \frac{1}{4!}\lambda_3 \phi_3^4\ .
\end{align}
In principle, higher dimensional operators could be added to this effective Lagrangian, though these turn up at one higher order in $y$ than we work, using the power counting prescription of equation \eqref{eq: power counting prescription} \cite{Braaten:1995cm, Niemi:2021qvp}.

At a first-order phase transition, the different terms in equation \eqref{eq: 3d Lagrangian} balance to yield two minima with a maximum between. For this to be possible, the different terms in the potential must be the same order of magnitude, such as $m_3^2 \phi_3^2 \sim \lambda_3 \phi_3^4$. Extending the power counting to the 3d couplings, $m_3^2 \sim m^2 \sim y^2 T^2$ and $\lambda_3 \sim \lambda T \sim y^2 T$, this relation leads to 
\begin{align}
    \Rightarrow \phi_3^2 &\sim T \, ,
\end{align}
so that relatively strong phase transitions are possible in this model. Note that the powers of temperature arise through first scaling the 3d scalar field to canonically normalise it, and then absorbing all explicit factors of $T$ into the parameters of the EFT \cite{Braaten:1995cm, Kajantie:1995dw}.

At high temperatures and at leading order (LO), the free energy density of the Yukawa theory is $f=-\pi^2 T^4/20$, independently of the background scalar field. For first-order phase transitions, it is the difference between phases $\Delta f$ which determines the dynamics. For homogeneous phases, the free energy density is equal to the thermal effective potential, and its expansion takes the form
\begin{align} \label{eq: y2 y5 count}
\frac{\Delta f}{T^4} &= \underbrace{c_2 y^2}_{\textup{1-loop}}
+ \underbrace{c_{3} y^3}_{\textup{1-loop}^+}
+ \underbrace{c_4 y^4}_{\textup{2-loop}^+}
+ \underbrace{c_{5} y^5}_{\textup{3-loop}^+}
+ \underbrace{c_{6} y^6}_{\textup{4-loop}^+}
+ \dots
\end{align}
where the superscript $^+$ indicates that these terms require resummation to compute. The coefficients $c_n$ are all of $\mathcal{O}(1)$ in our power counting, and are temperature dependent. In this work, we have evaluated the free energy density up to the $\mathcal{O}({y^5T^4})$ term, i.e.~calculating up to $c_5$. This is then compared to the result at leading order, with the free energy density at $\mathcal{O}({y^2T^4})$, i.e.~just $c_2$. For brevity, we will refer to results using the full calculation as $\mathcal{O}({y^5})$ and leading order results as $\mathcal{O}({y^2})$.

Loops of the hard energy scale $\pi T$ (i.e.~the dimensional reduction) yield an expansion in integer powers of $y^2$. The half-integer powers come from loops within the soft scale EFT, where the energy scale is $y T$ (or $\sqrt{\lambda}T$). The expansion parameter for the soft scale is $\alpha_3 \sim y$; see equation \eqref{eq: alpha3 expression}.

To derive the effective couplings of the EFT, we matched the off-shell correlation functions at soft external momenta: zero Matsubara modes with $p\sim y T$. We evaluated the connected, one-particle irreducible correlation functions, denoted by $\Gamma^{(k)}$ in the 4d theory and $\Gamma_3^{(k)}$ in the 3d effective theory. The calculations yield generic soft-scale observables accurate up to the $\mathcal{O}({y^5 T^4})$ term for the free energy density. These are given in appendix \ref{appendix: correlation functions}. By matching these physical observables, we arrive at expressions for the 3d effective parameters:
\begin{align}
\begin{split} \label{eq:dr result 1}
\phi_3^{2} ={} & \frac{\bar{\varphi}_0^2}{T}\left( 1 + \frac{4 y^2 }{3 (4 \pi)^2}\right)
\end{split}\\
\begin{split} \label{eq:dr result 2}
\sigma_3 ={} & \left(\frac{\bar{\sigma}}{\sqrt{T}} + \frac{\bar{g} T^{\frac{3}{2}}}{24} + \frac{\bar{y} \bar{m}_\psi T^{\frac{3}{2}}}{6} \right) \left(1 -\frac{2y^2}{3(4\pi)^2}\right) 
+ \frac{g \lambda T^{\frac{3}{2}}}{6 (4 \pi)^2}\left [ \log \left (\frac{\Lambda}{3T} \right ) - c \right ] + \frac{y^2g T^{\frac{3}{2}}\log 2}{2(4\pi)^2}
\end{split}\\
\begin{split} \label{eq:dr result 3}
m^2_3 ={} &
\left(\bar{m}^2 + \frac{\bar{\lambda} T^2}{24} + \frac{ \bar{y}^2 T^2}{6}\right)  \left(1 -\frac{4y^2}{3(4\pi)^2}\right)
+ \frac{\lambda^2 T^2}{6 (4 \pi)^2}\left [ \log\left (\frac{\Lambda}{3T} \right ) - c \right ]
 + \frac{y^2\lambda T^2\log 2}{2(4\pi)^2}
\end{split}\\
\begin{split} \label{eq:dr result 4}
g_3 ={} &  \sqrt{T} \bar{g} \left(1 - \frac{2 y^2}{(4 \pi)^2}\right)
\end{split}\\
\begin{split} \label{eq:dr result 5}
\lambda_3 ={} &  T \bar{\lambda} \left(1 - \frac{8 y^2}{3(4 \pi)^2}\right) \ ,
\end{split}
\end{align}

\noindent where couplings either do not have running at $\mathcal{O}({y^4})$, or are barred couplings defined to make transparent the renormalisation scale invariance at $\mathcal{O}({y^4})$:
\begin{align}
\bar{\chi} ={} & \chi - \frac{1}{2} \beta_{\chi}^\text{b} L_b(\Lambda) - \frac{1}{2} \beta_{\chi}^\text{f} L_f(\Lambda)
\end{align}

\noindent where $\beta_{\chi} = \beta_{\chi}^\text{b} + \beta_{\chi}^\text{f}$ denotes the beta functions, given in appendix \ref{appendix: zero temperature counterterms}, $L_b(\Lambda)$ and $L_f(\Lambda)$ are logarithms of the renormalisation scale, defined following ref.~\cite{Kajantie:1995dw} and given in appendix \ref{appendix: correlation functions}, $\gamma_E$ is the Euler–Mascheroni constant, and c is a constant arising from 2-loop sum-integrals, also given in appendix \ref{appendix: correlation functions}. Similarly to the barred couplings, we defined:
\begin{align}
\bar{\varphi}_0 =  \varphi_0 - \frac{1}{2} \gamma_\phi \varphi_0 L_f(\Lambda) \ ,
\end{align}

\noindent where $\gamma_\phi = - \frac{1}{\varphi_0} d \varphi_0 / d (\log \Lambda) = -2 y^2 / (4 \pi)^2$. All couplings and fields here are $\overline{\text{MS}}$ renormalised. These dimensional reduction relations extend the leading-order results of ref.~\cite{Gould:2021ccf}.

These expressions for the 3d effective parameters now enable perturbative analysis of the EFT to yield results applicable to the 4d Yukawa model. The scale $\Lambda$ can be replaced with two separate renormalisation scales, $\mu$ and $\mu_3$ which may be chosen independently \cite{Kajantie:1995dw}. The $\Lambda$ scales explicitly shown in equations \eqref{eq:dr result 2} and \eqref{eq:dr result 3} reproduce the renormalisation group running of the 3d EFT, allowing us to exchange for some new scale $\mu_3$ that can be chosen independently. Given the superrenormalisability of the 3d EFT, this running can be upgraded to be exact by replacing $g \lambda T^{3/2} \to g_3 \lambda_3$ and $\lambda^2 T^2 \to \lambda_3^2$ in front of the logarithms of $\Lambda$ in equations \eqref{eq:dr result 2} and \eqref{eq:dr result 3}. The remaining $\Lambda$ scale dependence carried implicitly in the 4d \MSbar\ parameters is then denoted as $\mu$ for clarity.

\section{Phase transitions and latent heat}

\subsection{Conditions for first-order phase transition}

The pattern of phase transitions in this 3d EFT has been studied in ref.~\cite{Gould:2021dzl}. The existence and order of any phase transitions can be determined as follows. One
 can find a basis in which $g_3(T) = 0$ for all $T$ by shifting $\phi_3 \rightarrow \phi_3 - g_3 / \lambda_3$. The bare potential then takes the following form:
\begin{equation} \label{eq:potential}
    V = \tilde{\sigma}_3 \phi_3 + \frac{1}{2} ( \tilde{m}_3^2 + \delta m_3^2 ) \phi_3^2 +\frac{1}{4!} \lambda_3 \phi_3^4 \ ,
\end{equation}
where we have defined
\begin{align} 
\tilde{\sigma}_3 &= \sigma_3 + \frac{g_3^3}{3 \lambda_3^2} - \frac{g_3 m_3^2}{\lambda_3} \ , &
\label{eq: eqn for r} \rr &= m_3^2 - \frac{g_3^2}{2 \lambda_3}\ .
\end{align}

The mass counterterm is given in equation \ref{eq:3d mass counterterm}, and the possible tadpole counterterm cancels. This reduces the theory to the $Z_2$-symmetric $\phi_3^4$ theory coupled to a finite $Z_2$-breaking external field $\tilde{\sigma}_3(T)$. A phase transition, if it exists, occurs at
\begin{equation} \label{eq:sigma_tilde_zero}
\tilde{\sigma}_3(\Tc) = 0\ ,
\end{equation}
where the symmetry is restored. This defines the critical temperature, $\Tc$. As a consequence of equation \eqref{eq:sigma_tilde_zero}, the properties of the phase transition, such as its order and strength, may depend only on $\tilde{m}_3^2$ and $\lambda_3$.%
\footnote{One may also include $\mu_3$ here. However, physical quantities are independent of $\mu_3$, and the dependence of intermediate quantities on $\mu_3$ is fixed by the superrenormalisability of the 3d theory.}
Further, by dimensional analysis, there can only be non-trivial dependence on the dimensionless ratio $\rr/\lambda_3^2$. Using mean field theory, one can see that for $\rr/\lambda_3^2 \ll -1$ there is a first-order phase transition, and for $\rr/\lambda_3^2 \gg 1$ there is a crossover. In between, the line of first-order phase transitions ends in a second-order phase transition at $\rr=\rstar$. The value of $\rstar$ has been determined using lattice Monte-Carlo simulations to be \cite{Sun:2002cc, Gould:2021dzl}
\begin{equation} \label{eq: eqn for r*}
\rstar = \left[0.0015249(48) + \frac{1}{6 (4 \pi)^2} \log \left( \frac{3 \mu_3}{\lambda_3} \right) \right] \lambda_3^2 \ ,
\end{equation}
with the number in parenthesis being the statistical uncertainty in the last digits, and the whole expression is evaluated at the critical temperature.

Due to the $Z_2$ symmetry at the critical temperature, $\phi_3 \to -\phi_3$, the condition $\tilde{\sigma}_3=0$ is exact, and is unmodified by loop corrections within the 3d EFT which can yield odd powers of $y$. Therefore, the perturbative expansion of $\Tc$ is one in powers of $y^2$, all odd powers of $y$ cancelling due to the symmetry,
\begin{align} \label{eq:Tc expansion}
   \Tc = \Tc^{(0)}(1 + a_2 y^2 + a_4 y^4 + \dots).
\end{align}
The power counting of other couplings and parameters are related to $y$ using the prescription of equation \eqref{eq: power counting prescription}. The LO result $\Tc^{(0)}$ can be computed with an unresummed 1-loop calculation, while the next-to-leading order correction $a_2$ requires 2-loop dimensional reduction to be carried out with the free energy evaluated to at least $\mathcal{O}({y^4T^4})$. The coefficients $a_n$ can be obtained analytically by solving equation \eqref{eq:sigma_tilde_zero} in a strict expansion in powers of $y$.

The order of the transition is determined by the sign of $\drstar$ at the critical temperature. $\rr$ is determined by purely hard-scale physics and $\rstar$ is determined by purely soft-scale physics. However, the soft physics is known from lattice simulations, so the only remaining uncertainties in $\drstar$ come from the hard scale.

\subsection{Phase transitions for Yukawa theory} \label{subsection: PTs outline}

For a given parameter point in our Yukawa model, the critical temperature (if it exists) and the order of any phase transition can be determined by combining equations \eqref{eq:sigma_tilde_zero} and \eqref{eq: eqn for r*} with the expressions for the 3d effective parameters, equations \eqref{eq:dr result 1} to~\eqref{eq:dr result 5}.

Using the dimensional reduction relations at leading order, equation \eqref{eq:sigma_tilde_zero} for the critical temperature has a simple analytic solution, about which subleading corrections can be obtained by a strict perturbative expansion. Our higher order results include the subleading term suppressed relatively by $y^2$; see equation \eqref{eq:Tc expansion}. In what follows, we give explicit analytic results for the critical temperature to this order.

To simplify the results, we introduce 4d tilded parameters, defined in analogy with those in 3d by transforming $\phi \to \phi - g/\lambda$ to eliminate the cubic scalar term.
In terms of the (untilded) parameters of the original basis, the (tilded) parameters of the new basis read
\begin{align} \label{eq:4d tilde}
    \tilde{\sigma} &= \sigma - \frac{g m^2}{\lambda} + \frac{g^3}{3\lambda^2}, &
    \tilde{m}^2 &= m^2 - \frac{g^2}{2\lambda}, &
    \tilde{g} &= 0, &
    \tilde{m}_\psi &= m_\psi - \frac{gy}{\lambda},
\end{align}
and $y$ and $\lambda$ are unchanged.

The coefficient of the symmetry-breaking term in the 3d EFT then reads
\begin{align} \label{eq:sigma3 tilde}
    \tilde{\sigma}_3(T) = \frac{1}{\sqrt{T}}\bigg\{&
    \tilde{\sigma} + \frac{1}{6}y \tilde{m}_\psi T^2
    + \frac{1}{(4\pi)^2}\bigg[
   4 y \tilde{m}_{\psi }^3 L_f
   -y^2 \tilde{\sigma}  \left(L_f +\frac{2}{3}\right)
   -\frac{24 y^3 \tilde{m}^2 \tilde{m}_{\psi }}{\lambda} L_f
   \nonumber\\
   &\qquad
   +\left(
        y^3 \tilde{m}_{\psi}\left(\frac{1}{3} L_b - L_f -\frac{1}{9}\right)
        -\frac{4 y^5 \tilde{m}_{\psi }}{\lambda} L_f
   \right) T^2 
    \bigg]
    + \mathcal{O}(y^4\tilde{\sigma})
    \bigg\}.
\end{align}

Solving $\tilde{\sigma}_3(\Tc) = 0$ in a strict expansion, using the free energy to $\mathcal{O}({y^5T^4})$, we find the critical temperature,
\begin{align} \label{eq: TC}
    \Tc^2 =
    -\frac{6 \tilde{\sigma }}{y \tilde{m}_{\psi }}\left[1
    + \frac{1}{(4\pi)^2}\left(
    \frac{4 y \tilde{m}_{\psi }^3}{\tilde{\sigma}} L_f
    + y^2 (5 L_f - 2 L_b)
    + \frac{24 y^4}{\lambda} \left(1 - \frac{\tilde{m}^2 \tilde{m}_{\psi}}{y\tilde{\sigma}}\right) L_f
    \right)
    + \mathcal{O}(y^4)\right].
\end{align}

At leading order, this can be expressed in terms of the original 4d parameters as 
\begin{align} \label{eq: LO TC}
\Tc^{(0)} = \sqrt{
\frac{2(g^3-3 g \lambda m^2+3\lambda ^2\sigma)}{y \lambda \left(y g-\lambda m_{\psi }\right)}
}  \ .
\end{align}

For $y=0$, the scalar and fermion decouple, and $\sqrt{T} \tilde{\sigma}_3$ becomes independent of $T$ for any choice $\mu \propto T$, so $\tilde{\sigma}_3$ cannot change sign for the scalar-only theory, at least up to this order. As a consequence the only possible phase transition for the scalar-only theory is of second order, where $\tilde{\sigma}_3=0$ for all temperatures and the transition takes place when $\drstar$ goes through zero \cite{Sun:2002cc}. Similarly, $\tilde{\sigma}_3$ cannot change sign in the $Z_2$ limit of this model, where $\tilde{\sigma}=\tilde{m}_\psi=0$, but $y\neq 0$.

If $\Tc^2$ is negative, then there is no phase transition. If it is positive, equations \eqref{eq: eqn for r} and \eqref{eq: eqn for r*} can be used to determine $\drstar$ and hence the order of the phase transition. The 3d effective mass parameter has a perturbative expansion in integer powers of $y^2$, which, when evaluated at the critical temperature, reads
\begin{align} \label{eq: m3sqtilde}
    \rr &=
    \tilde{m}^2-\frac{\tilde{\sigma }}{4 y \tilde{m}_{\psi }} \left(\lambda +4 y^2\right)
    + \frac{1}{(4\pi)^2} \bigg\{
    -\frac{4}{3} y^2 \tilde{m}^2
    -\frac{1}{2} \lambda  \tilde{m}^2 L_b
    + \left(4y^2 +24 \frac{y^4}{\lambda}\right) \tilde{m}^2 L_f 
    \nonumber \\
    &\qquad\qquad
   +\left(8 y^2-\lambda \right) \tilde{m}_{\psi }^2 L_f
    + \frac{\tilde{\sigma }}{y \tilde{m}_\psi}
    \bigg[
        \frac{4}{3} y^4
        + \left(\frac{1}{3} - 3 \log 2\right) \lambda y^2
        +  \left(\frac{3}{8} \lambda^2 + \frac{1}{2} \lambda y^2\right) L_b
        \nonumber \\
        &\qquad\qquad\qquad
        -\left(\frac{1}{4}\lambda  y^2 + 10 y^4 + 24 \frac{y^6}{\lambda }\right) L_f
        - \lambda ^2 \left(
            \log \left(\frac{B}{3}\right) - c
        \right)
    \bigg]
    \bigg\}
    + \mathcal{O}(\tilde{m}^2 y^4) \,,
\end{align}
where we have chosen both renormalisation scales proportional to the temperature, with $\mu = A \pi T$, and $\mu_3 = B T$.

Renormalisation scale dependence can be used to estimate the size of missing higher order terms, and so to give an intrinsic measure of the theoretical uncertainty in a prediction. In a strict expansion of a physical quantity, any renormalisation scale dependence cancels exactly, order by order. However, by solving for the running of the parameters to a higher order than the underlying calculation, a formally higher order renormalisation scale dependence can be introduced.

In practice, we have solved the 1-loop renormalisation group equations perturbatively, working to one higher order than the thermodynamic calculation; see appendix \ref{appendix: zero temperature counterterms} for the beta functions. That is, denoting a 4d \MSbar\ parameter by $\kappa$, we have solved for the running up to and including the $\mathcal{O}(\kappa y^2)$ term for our LO thermodynamics, and up to the $\mathcal{O}(\kappa y^4)$ term for our higher order thermodynamics. As before, $y$ here is used as a shorthand to include terms of equivalent sizes according to the power counting scheme in equation \eqref{eq: power counting prescription}. Writing a given \MSbar\ parameter $\kappa$ as a function of $t=\log\left(\mu'/\mu\right)$, the higher order perturbative solution to the running is
\begin{align}
    \kappa(t) \approx \kappa(0) + \beta_\kappa(0)t + \frac{1}{2}\sum_{a} \frac{\partial \beta_{\kappa}(0)}{\partial \kappa_a(0)} \beta_{\kappa_a}(0) t^2,
\end{align}
where $\kappa_a$ runs over all the parameters in the model. Note that while 2-loop contributions to the beta functions would contribute at $\mathcal{O}(\kappa y^4)$, we do not include them, as we are merely using the running to estimate uncertainties.
These running couplings were then included in expressions for the thermodynamic quantities, such as eqs.~\eqref{eq: TC} and \eqref{eq: m3sqtilde}. By then varying the renormalisation scale over a range, we can obtain a measure of the intrinsic uncertainty of the predictions. We chose the renormalisation scales as $\mu=A \pi \Tc$, where $A$ varies over 7 equidistant values from $1/2$ to $2$. This range follows standard practice in the field \cite{Laine:2017hdk, Gould:2021oba}. The arithmetic mean of the results is taken as the predicted physical quantity, and the minimum and maximum values yield the range of uncertainty. For simplicity, we fix $B=1$ throughout.

\subsection{Scan of parameter space} \label{subsection: scan}

A global study of the parameter space was carried out. Without loss of generality, units were chosen such that $m^2=1$, and $\sigma$ was set to zero (this can be fixed by a shift $\phi \to \phi + \text{constant}$). The distributions of the other parameters $y^2, g^2, \lambda \text{ and } m_\psi$ were chosen according to
\begin{align}
    y^2,\ \lambda,\ g^2 &\in 10^{\mathcal{U}(-2,~0.5)}, &
    m_\psi &\in \mathcal{U}(0, 3),
\end{align}
where $\mathcal{U}(a,b)$ is the uniform distribution on the interval $[a,b)$, and the signs of $y$ and $g$ were chosen randomly and with equal probability.
Perturbativity at zero temperature requires that all three couplings are small compared with $\sim 8\pi$ (see e.g.~\cite{Allwicher:2021rtd}), hence the maximum of the range, and a log distribution was chosen to explore a wide range of possible ratios of couplings.
A linear distribution was chosen for the fermion mass parameter because the boundaries between different behaviours occur at $\mathcal{O}(1)$ values of $m_\psi$, and no special behaviours are expected (or observed) in the $m_\psi \to 0$ limit. Further, large values of $m_\psi$ are more likely to invalidate our high-temperature approximation.

If treated as physical parameters, to be consistent with the accuracy of the rest of our calculations, the values generated should be matched to \MSbar\ parameters at zero temperature at 1-loop order \cite{Kajantie:1995dw}. However, as we are primarily interested in the perturbative convergence of thermodynamic predictions, rather than interpreting the predictions in a cosmological context, we neglect this zero-temperature matching between \MSbar\ and physical parameters, instead treating this as a toy model. We therefore carry out a different scan, directly over $\overline{\text{MS}}$ parameter space at the input renormalisation scale $\mu_0 = 1$, carrying over their values directly to the phase transition evaluation.

\begin{figure}[ht]
\centering
\includegraphics[width=0.93\textwidth]{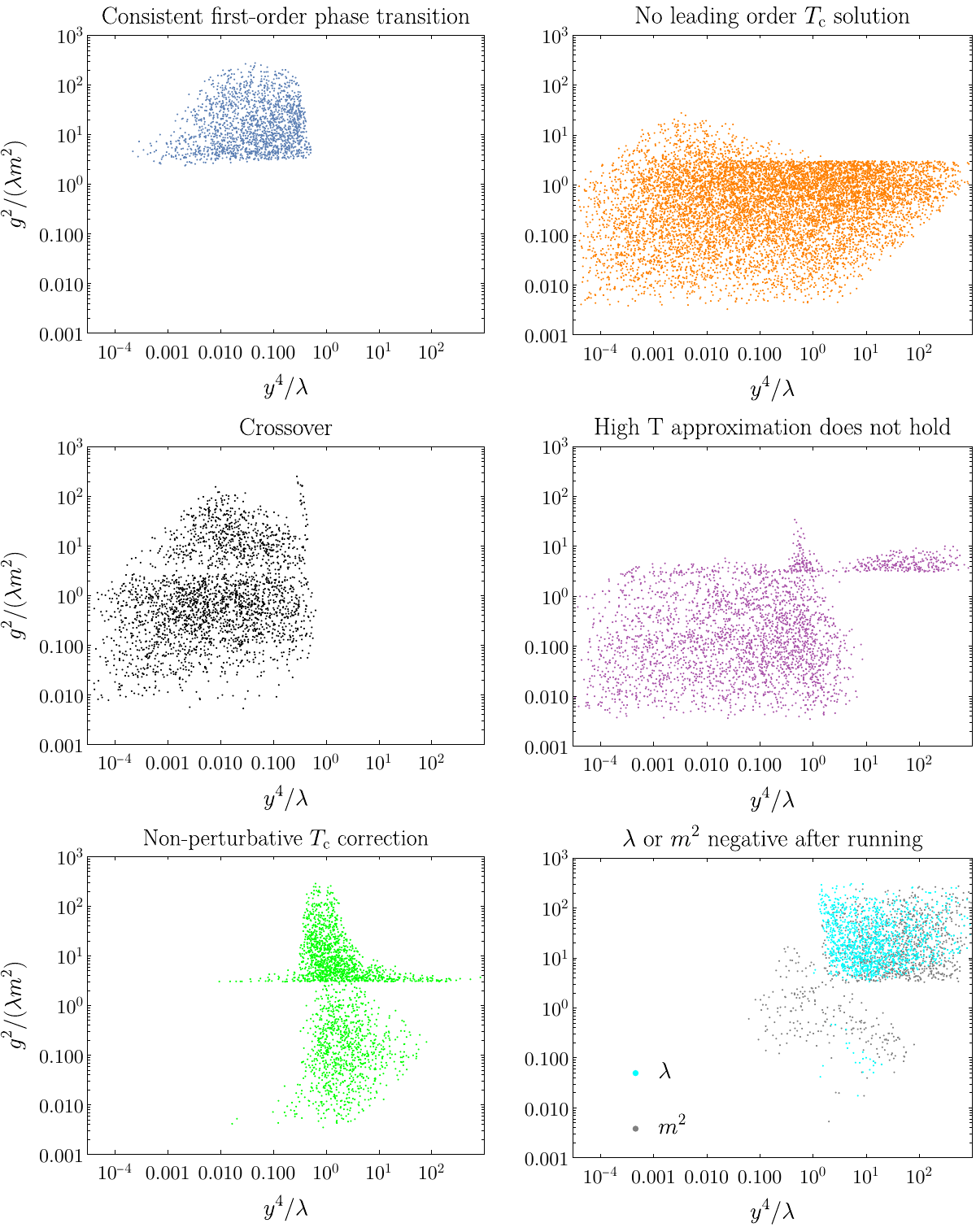}
\caption{Predicted nature of scalar singlet Yukawa theories studied with free energy evaluated to $\mathcal{O}({y^5T^4})$ for our scan of parameters. First-order phase transitions were found to be restricted to parameter space where $\frac{g^2}{\lambda m^2} \gtrsim 2$ and $\frac{y^4}{\lambda} \lesssim 0.5$. There are two additional directions in parameter space not shown here, $\lambda$ and $m_\psi/m$.
} 
\label{fig:Phase Transitions}
\end{figure}

Fig.~\ref{fig:Phase Transitions} shows the phase transition evaluation of the Yukawa theory, using the full $\mathcal{O}({y^5})$ calculation as per equation \eqref{eq: y2 y5 count}, for a scan of 20,000 parameter points using the distributions summarised above. The axes are chosen as combinations of parameters which bound the first-order phase transitions identified. We find that first-order phase transitions are limited to parameter space where $\frac{g^2}{\lambda} \gtrsim 2 m^2$. At leading order, this condition follows from the requirement $\drstar < 0$ for first-order phase transitions, which can be written as
\begin{equation}
\frac{g^2}{2 \lambda} > m^2 + \frac{1}{24} (\lambda + 4 y^2) \Tc^2 \ ,
\end{equation}
with the second term positive definite.

Parameter points are also shown which have no leading order solution for the critical temperature (in orange), or are predicted to be crossovers (in black). Following this are parameter points, shown in purple, where the high temperature approximation does not hold, defined as $\Tc < 2\cdot \text{max}(m, m_\psi)$. This precise condition is motivated by the convergence of the high-temperature expansion for the 1-loop thermal functions \cite{Laine:2016hma}.

It can also be seen from fig.~\ref{fig:Phase Transitions} that for values of $\frac{y^4}{\lambda} \gtrsim 0.5$, our power counting assumptions can begin to break down. For the points in green, the correction to the critical temperature from formally subleading terms in the $\mathcal{O}({y^5T^4})$ free energy density and from running is larger than the leading order value.
We have denoted as non-perturbative such points which satisfy
\begin{equation}
     |\Tc(\mu) - T_\text{c}^\text{(0)}| > T_\text{c}^\text{(0)} \, ,
\end{equation}
for some renormalisation scale $\mu$ in the range considered, where $\Tc(\mu)$ is the critical temperature computed with the $\mathcal{O}({y^5T^4})$ free energy density and $\mathcal{O}({y^6})$ running, and $\Tc^\text{(0)}$ is the critical temperature evaluated at $\mathcal{O}({y^2})$ with no running. At leading order, an analogous non-perturbative condition can be defined, with $\Tc(\mu)$ now the critical temperature computed to $\mathcal{O}({y^2})$ with running at $\mathcal{O}({y^4})$.

For the parameter points in cyan, $\lambda$ is negative after running, which can be interpreted a breakdown of the model. Finally, grey points have negative $m^2$ after running, which indicates large corrections from perturbative renormalisation group running.

The theories identified as first-order phase transitions were required to be consistently evaluated as such across the range of renormalisation scales $\mu=A \pi \Tc$, where $A$ varies from $1/2$ to $2$. This is further discussed in section \ref{subsection: reduced errors}.

As there are only four independent parameters, it has been feasible to attain a reasonably dense scan of the relevant parameter space with a simple random scan, and to illustrate features of the results with projections onto a plane in parameter space. While the distribution of our results inevitably depends on the details of our scan choice, all our calculations were carried out on the same set of parameter points, allowing us to reveal differences in predictions based on the $\mathcal{O}(y^2)$ and $\mathcal{O}(y^5)$ approximations. In theories with higher dimensional parameter spaces, we would expect adaptive search algorithms to become necessary for efficient study of the phase transitions of a model \cite{AbdusSalam:2020rdj}.

\subsection{Latent heat evaluation}

After determining the type of phase transition at a parameter point, if any, attention is next turned to the strength of the phase transition in terms of the latent heat. Strong phase transitions are of particular interest for detectable gravitational wave signals.

The latent heat, $L$, of a first-order phase transition can be determined by the following thermodynamic relation evaluated at the critical temperature:
\begin{equation}
L = - \frac{\partial \Delta f}{\partial \log T}\bigg|_{\Tc} \ ,
\end{equation}
where $f$ is the free energy density of the full 4d theory and $\Delta f \equiv f_{+} - f_{-}$, where $f_{+}$ and $f_{-}$ are the free energy densities in the higher and lower temperature phases respectively. This can be expressed in terms of the effective parameters of the 3d EFT as:
\begin{equation}
L = - \Tc \frac{\partial \tilde{\sigma}_3}{\partial \log T}\bigg|_{\Tc} \Delta \langle \bar{\phi}_3 \rangle \ ,
\end{equation}
where $\tilde{\sigma}_3$ is given in equation \eqref{eq:potential} and $\langle \bar{\phi}_3 \rangle$ is the linear field condensate.
The definition of the former is structurally akin to a beta function.
It depends only on the UV physics of the hard scale, and can be read off from equation \eqref{eq:sigma3 tilde}.
The field condensate depends on the IR physics of the soft-scale EFT. At leading order, it is related to:
\begin{equation}
 \Delta \langle \bar{\phi}_3 \rangle  = 2 v_0 \ ,
\end{equation}
where
\begin{equation}
v_0^2 = \frac{-6 \rr}{\lambda_3} \ 
\end{equation}
is the field value of the tree level minima about the Z\textsubscript{2}-symmetric origin.

The jump in the linear condensate has been evaluated to three loops within the EFT, and is given by \cite{Gould:2021dzl}:
\begin{equation}
\Delta \langle \bar{\phi}_3 \rangle = v_0 \left\{2 + \sqrt{3} \alpha_3 + \left[\frac{1}{2} + \log \left(\frac{\mu_3}{\sqrt{3 \lambda_3} v_0}\right)\right] \alpha_3^2 - 1.15232\ \alpha_3^3 + \mathcal{O}(\alpha_3^4) \right\} \ ,
\end{equation}
where $\mu_3$ is the 3d EFT $\overline{\text{MS}}$ renormalisation scale, and we have introduced the loop-expansion parameter $\alpha_3$:
\begin{align} \label{eq: alpha3 expression}
\alpha_3 \equiv  \frac{\sqrt{\lambda_3}}{4 \pi |v_0|}.
\end{align}
Note that $\lambda_3 \sim y^2 T$, while $v_0^2 \sim T$ so that $\alpha_3 \sim y$.

\begin{figure}[h]
\centering
\includegraphics[width=0.7\textwidth]{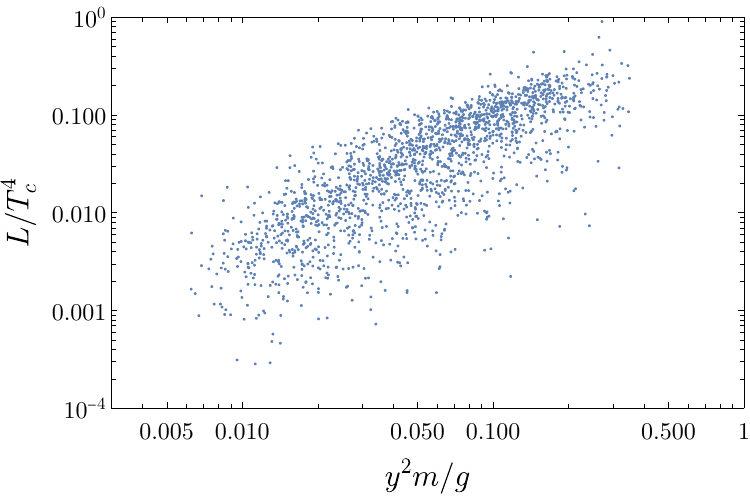}\hspace*{1.5cm}
\caption{The strength of the phase transition, represented here by the dimensionless $L/\Tc^4$, is plotted for all the first-order phase transitions in the parameter space studied within our scan, evaluated at $\mathcal{O}({y^5})$. It is found to be positively correlated with the ratio $y^2m/g$.} 
\label{fig:Latent Heat}
\end{figure}

$L/\Tc^4$ is a dimensionless measure of the latent heat, and therefore of the strength of the first-order phase transition. Similarly to the critical temperature, we evaluated this using free energy evaluated to both $\mathcal{O}({y^2T^4})$ and $\mathcal{O}({y^5T^4})$. Parameters were run to higher orders, and varied over a range of renormalisation scales, to measure the uncertainty in the predictions.

Fig.~\ref{fig:Latent Heat} plots $L/\Tc^4$ for \textit{consistent first-order phase transitions}, with a maximum value of $\sim0.5$ found in our scan at $\mathcal{O}(y^5)$. The latent heat is plotted against $y^2 m/g$, with a strong positive correlation between the two. Since theories with $y^2 m/g \gtrsim 0.5$ begin to be non-perturbative in our analysis (Fig.~\ref{fig:Phase Transitions}), a follow up investigation could identify additional, stronger, first-order phase transitions at larger values, for which a different power counting may be necessary, as $y^2 m/g \sim y$ is no longer small.

\section{Comparing results to leading order}

The calculations in this work have been carried out to three powers of $y$ beyond leading order, as set out in equations \eqref{eq: power counting prescription} and \eqref{eq: y2 y5 count}. This section will investigate the size and properties of the sub-leading corrections in the context of the properties of first-order phase transitions.

A number of works have found large discrepancies between calculations of phase transition parameters at low perturbative orders. For example, in the $Z_2$-symmetric limit of the real-scalar-extended SM, refs.~\cite{Carena:2019une, Ellis:2022lft} have found huge discrepancies between computations of the effective potential at order $\mathcal{O}(y^2T^4)$ and $\mathcal{O}(y^3T^4)$. Ref.~\cite{Gould:2021oba} found significantly reduced uncertainties in this model at order $\mathcal{O}(y^4T^4)$. Around the critical temperature, properties of the phase transitions are highly sensitive to additional sub-leading terms, contributing to large corrections to the properties of expected gravitational waves.

\subsection{Additional strong phase transitions}
\begin{figure}[ht]
\centering
\includegraphics[width=0.90\textwidth]{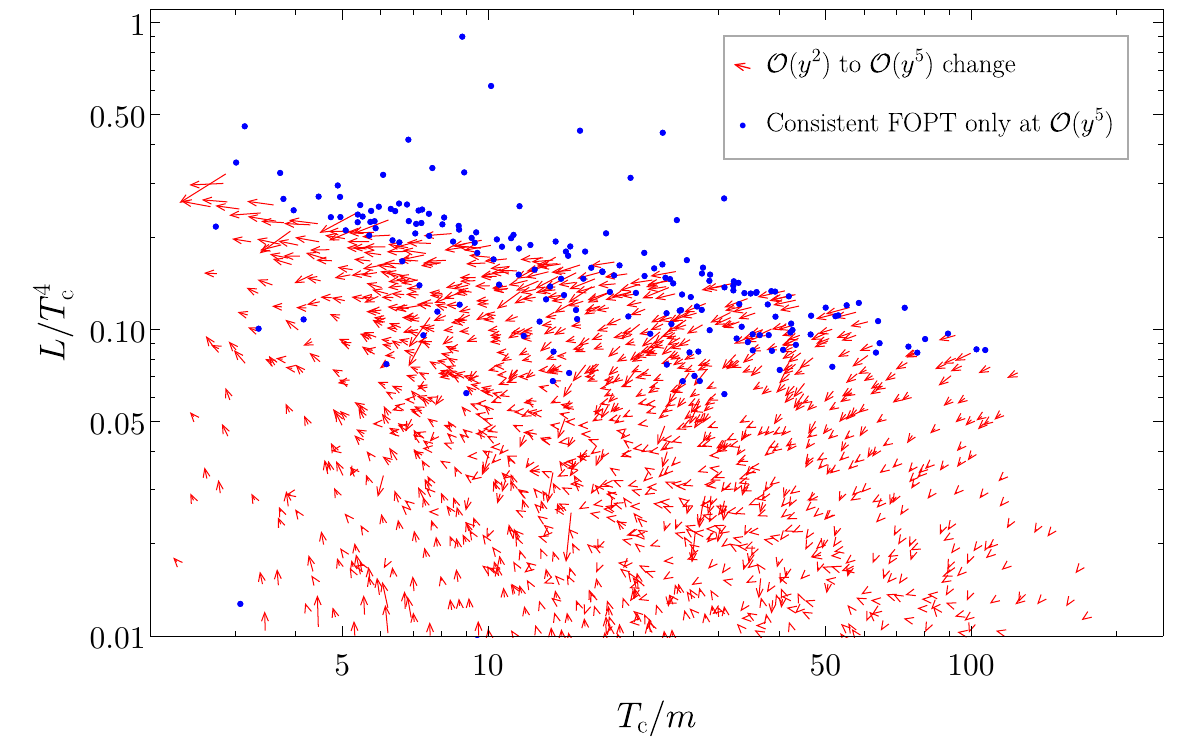}\hspace*{0.5cm}
\caption{The blue points show theories which are only \textit{consistent first-order phase transitions} (i.e.~valid across renormalisation scales) at $\mathcal{O}({y^5})$ and not at $\mathcal{O}({y^2})$, using power counting as per eqs.~\eqref{eq: power counting prescription} and \eqref{eq: y2 y5 count}. For \textit{consistent first-order phase transitions} at both $\mathcal{O}({y^2})$ and $\mathcal{O}({y^5})$, the changes due to loop corrections for the latent heat and critical temperatures are plotted as red arrows. There are no parameter points that are \textit{consistent first-order phase transitions} at leading order but not at $\mathcal{O}({y^5})$.} 
\label{fig:TCLH_LO_NLO}
\end{figure}

Across different renormalisation scales, there may not be only quantitative differences in the values of properties such as latent heat, but also qualitative differences in the predicted nature of the phase transition. This can mean for example that a particularly theory has a first-order phase transition at one renormalisation scale, but becomes a crossover when evaluated at another scale. 

\begin{figure}[t]
\centering
\includegraphics[width=0.90\textwidth]{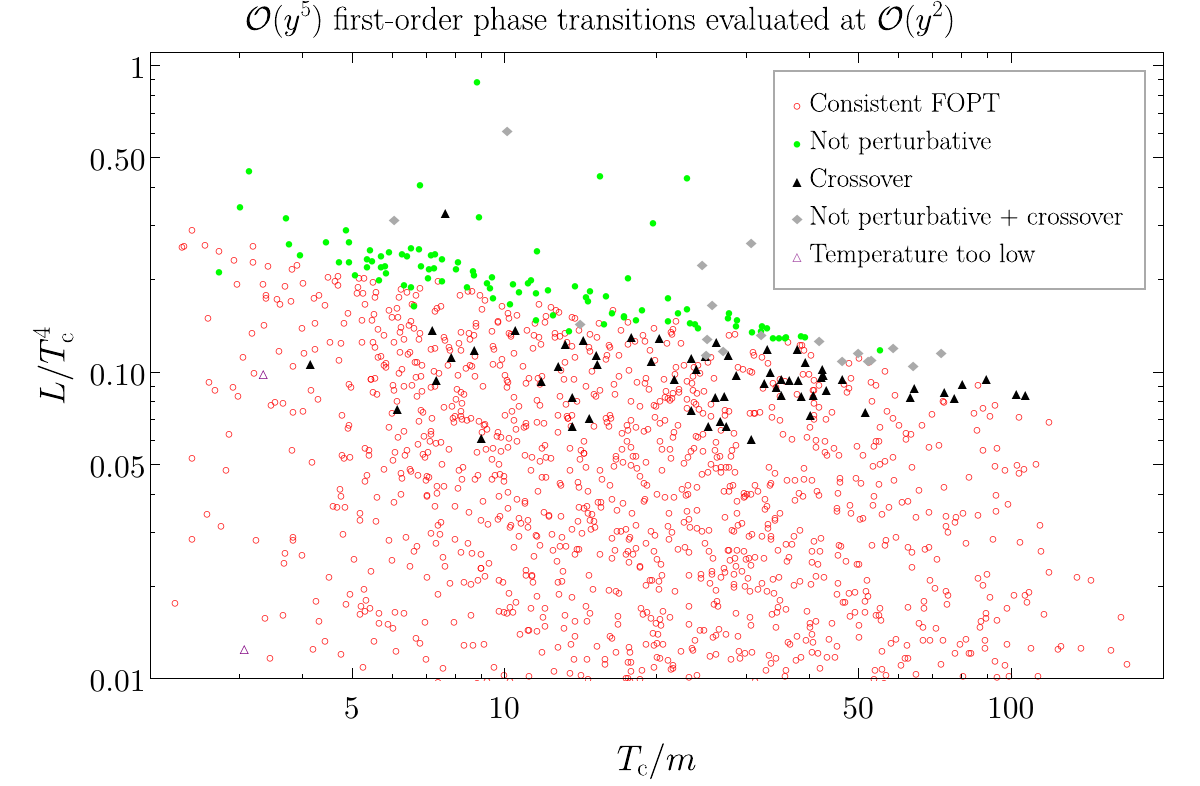}\hspace*{0.5cm}
\caption{
The leading order calculation failed to identify the majority of the strongest first-order phase transitions found at $\mathcal{O}({y^5})$. At one or more renormalisation scales, many of these were found to be not perturbative at leading order --- where contributions from the next order are larger than the leading order value of $\Tc$. Other strong phase transitions were categorised as crossovers at leading order.}
\label{fig:TCLH_LO_NLO_errors}
\end{figure}

This has been treated in ref.~\cite{Papaefstathiou:2021glr} through a characterisation from ``ultra-conservative" to ``liberal", based on the consistency of results with respect to scale. Here we adopt an approach that would be considered ``conservative" in this sense, where a theory associated with a set of parameters is considered to be a \textit{consistent first-order phase transition} if it is true across
the range of renormalisation scales $\mu=A \pi \Tc$, where $A$ varies over 7 equidistant values from $1/2$ to $2$. The theories shown in blue in fig.~\ref{fig:Phase Transitions} are all \textit{consistent first-order phase transitions} in this sense, whereas points in other categorisations were evaluated to have the corresponding prediction (e.g.~crossover) at one or more of the renormalisation scales considered.

It can be seen from figure \ref{fig:TCLH_LO_NLO} that a large number of parameter points are only found to be \textit{consistent first-order phase transitions} with the higher order $\mathcal{O}({y^5})$ calculation, and not when evaluated at leading order, $\mathcal{O}({y^2})$. The higher loop order calculation significantly changes the landscape of strong first-order phase transitions. The additional first-order phase transitions at $\mathcal{O}({y^5})$ in our scan correspond to a 57\% increase in FOPTs with $L/\Tc^4 > 0.1$ (from 229 to 359), or a 270\% increase in FOPTs with $L/\Tc^4 > 0.2$ (from 19 to 71).

Figure \ref{fig:TCLH_LO_NLO_errors} plots the same points, now also indicating the nature of the theories which are consistent first-order phase transitions at $\mathcal{O}({y^5})$, but not at $\mathcal{O}({y^2})$, for at least one renormalisation scale. Over half of these are due to non-perturbative corrections when evaluated at leading order, with running at $\mathcal{O}({y^4})$, as defined in section \ref{subsection: scan}.

\begin{figure}[h]
\captionsetup[subfigure]{margin={-0.6cm,0cm}}
\subfloat[\label{fig:benchmarks crossover to fopt}]{%
  \hspace*{0.4cm}\includegraphics[trim={2.9cm 7.4cm 5.3cm 2.5cm},clip,width=0.8\columnwidth, center]{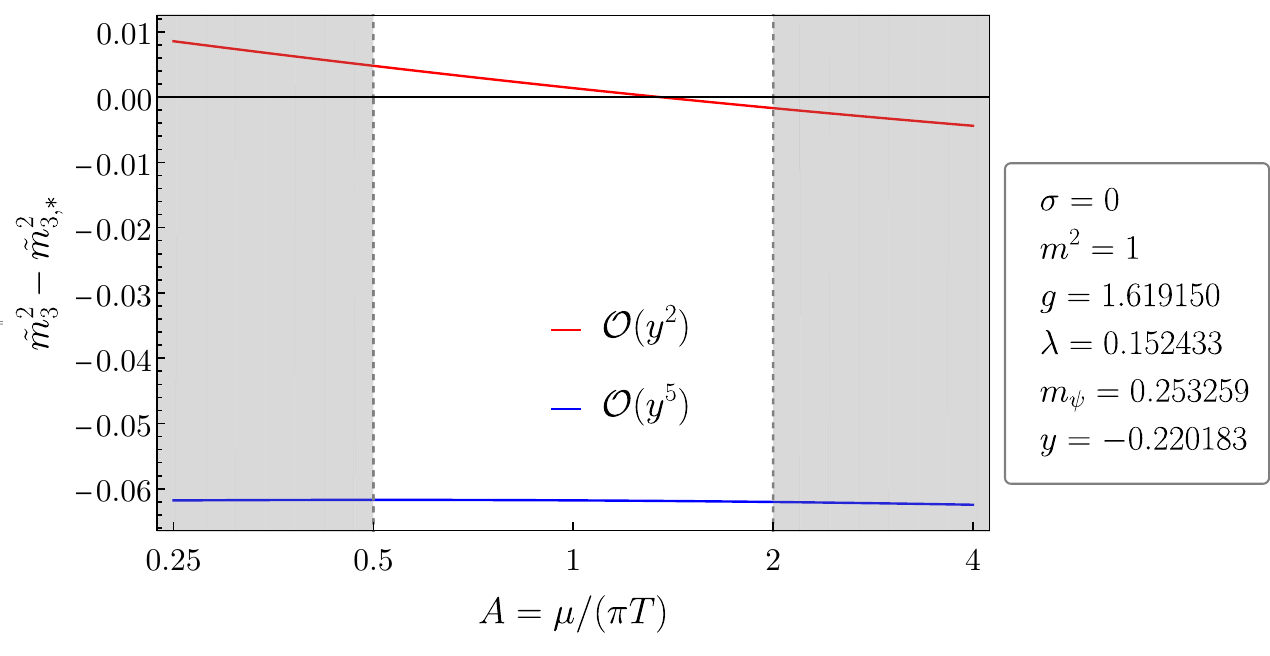}%
} 

\bigskip

\subfloat[\label{fig:benchmarks crossover to crossover}]{%
  \hspace*{0.4cm}\includegraphics[trim={2.9cm 7.4cm 5.3cm 2.5cm},clip,width=0.8\columnwidth, center]{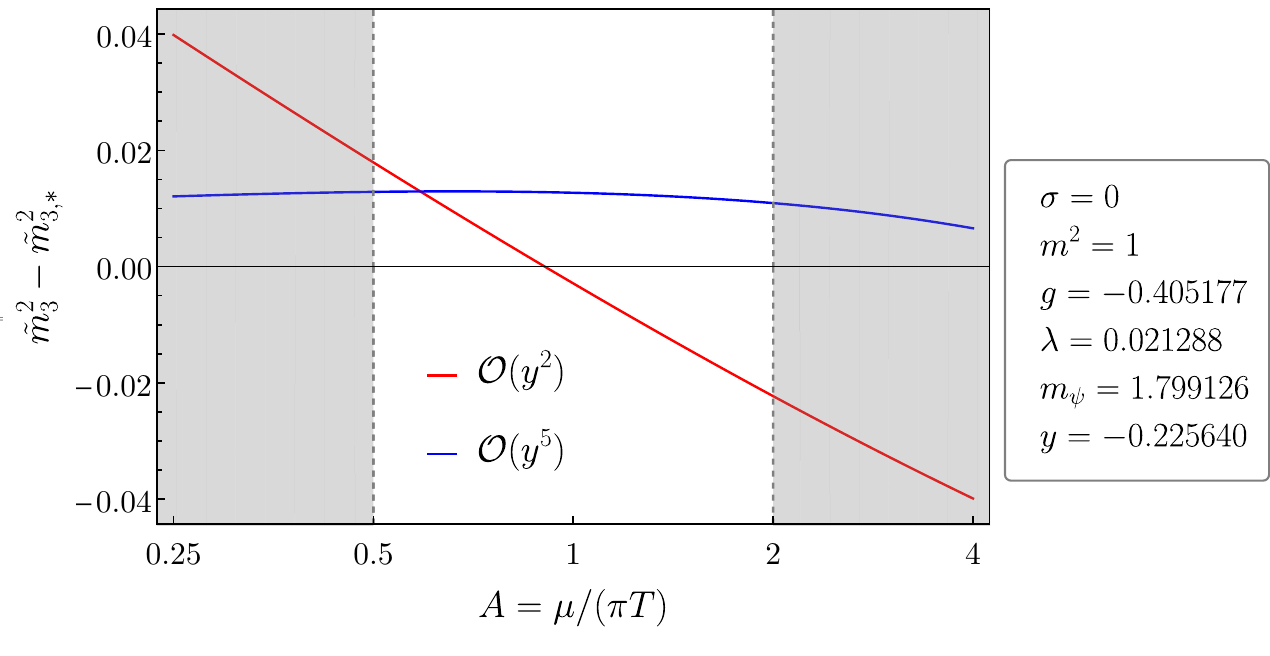}%
} 
\caption{Two parameters points which at $\mathcal{O}({y^2})$ are either crossovers or FOPTs, depending on renormalisation scale choice, are found at $\mathcal{O}({y^5})$ to be consistently FOPT(a) and crossover(b) irrespective of renormalisation scale.}
\label{fig:benchmarks sometimes crossovers at y2}
\end{figure}

Most of the remainder are predicted to be crossovers ($\drstar > 0$) for at least some part of the renormalisation scale range at leading order. Two example parameter points are shown in figure \ref{fig:benchmarks sometimes crossovers at y2}. In both cases, the $\mathcal{O}({y^2})$ evaluation indicates either a crossover ($\drstar > 0$) or a first-order phase transition ($\drstar < 0$), depending on the renormalisation scale used. Therefore, neither point is considered to be a \textit{consistent first-order phase transition} at leading order. Using the $\mathcal{O}({y^5})$ calculation, the parameter point in figure \ref{fig:benchmarks crossover to fopt} is confirmed to be a first-order phase transition, whereas the example in figure \ref{fig:benchmarks crossover to crossover} is found to be consistently a crossover across the renormalisation scales considered. If their natures were determined using only the $\mathcal{O}({y^2})$ calculation at a single renormalisation scale $\mu = \pi T$, the parameter point in figure \ref{fig:benchmarks crossover to fopt} would be determined as a crossover, and the point in figure \ref{fig:benchmarks crossover to crossover} a first-order phase transition, both opposites of the conclusions of the $\mathcal{O}({y^5})$ calculation.

\subsection{Movement of phase transitions}

Figure \ref{fig:TCLH_LO_NLO} also shows the change in latent heat and critical temperature for all \textit{consistent first-order phase transitions} identified at both leading order and $\mathcal{O}({y^5})$. The red arrows visualise distinctive flows of the higher order contributions in the latent heat and critical temperature plane for our parameter space scan. The latent heat corrections are generally positive, for phase transitions at lower critical temperature, and negative for those at higher critical temperature. The critical temperature corrections are generally negative. This pattern of flow means that a given set of theories occupying some parameter space, is transported as a whole in the latent heat and critical temperature plane, as a result of including higher order contributions. In other words, \textit{the blob moves}, at least in the context of our scan.

\begin{figure}[h]
\centering
\includegraphics[width=0.80\textwidth]{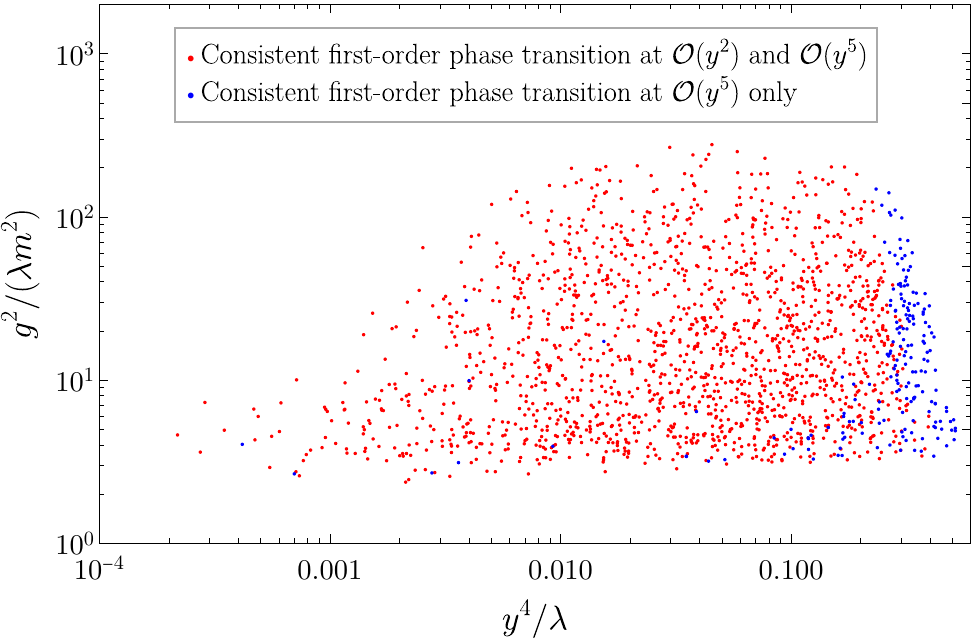}\hspace*{0.8cm}
\caption{From $\mathcal{O}({y^2})$ to $\mathcal{O}({y^5})$, \textit{consistent first-order phase transitions} expand in our scan's theory parameter space to include points with larger Yukawa couplings relative to $\lambda$.}
\label{fig:blob_param}
\end{figure}

As well as in the space of the physics predictions of the theory, it is also possible to study the transport of predictions in parameter space when higher order contributions are added. Figure \ref{fig:blob_param} plots theories predicted to be first-order phase transitions at $\mathcal{O}({y^2})$ and $\mathcal{O}({y^5})$, with power counting as before per eqs.~\eqref{eq: power counting prescription} and \eqref{eq: y2 y5 count}, against combinations of parameters $y^4/\lambda$ and $g^2/(\lambda m^2)$. These are the same combinations as fig.~\ref{fig:Phase Transitions}, where they are found to be correlated with the type of phase transition predicted.
The figure shows that, for our scan, within the higher order calculation the set of consistent FOPTs extends in the direction of larger $y^4/\lambda$, shown in blue. Therefore, in parameter space, the blob grows, but does not move, in the context of our scan.

At larger still values $y^4/\lambda \sim \pi^2$, 1-loop contributions from the Yukawa interaction compete with the tree-level scalar self-coupling in the zero-temperature effective potential. In this case a new set of power counting relations, and consequently resummations, are required for the perturbative description \cite{Coleman:1973jx, Weinberg:1992ds, Kierkla:2022odc}.

\subsection{Reduced errors} \label{subsection: reduced errors}

\begin{figure}[h]
\centering
\includegraphics[width=0.65\textwidth]{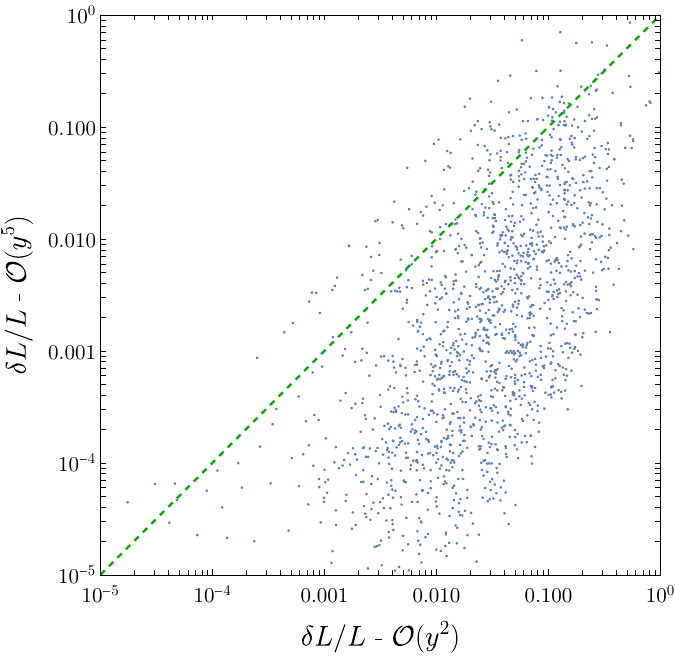}
\caption{There is generally a reduction in error in predictions of the latent heat at $\mathcal{O}({y^5})$ compared to $\mathcal{O}({y^2})$, in our scan, for theories that are \textit{consistent first-order phase transitions} at both orders.}
\label{fig:TCLH_LO_NLO_errors_compare}
\end{figure}

\begin{figure}[ht]
\centering
\includegraphics[width=0.9\textwidth]{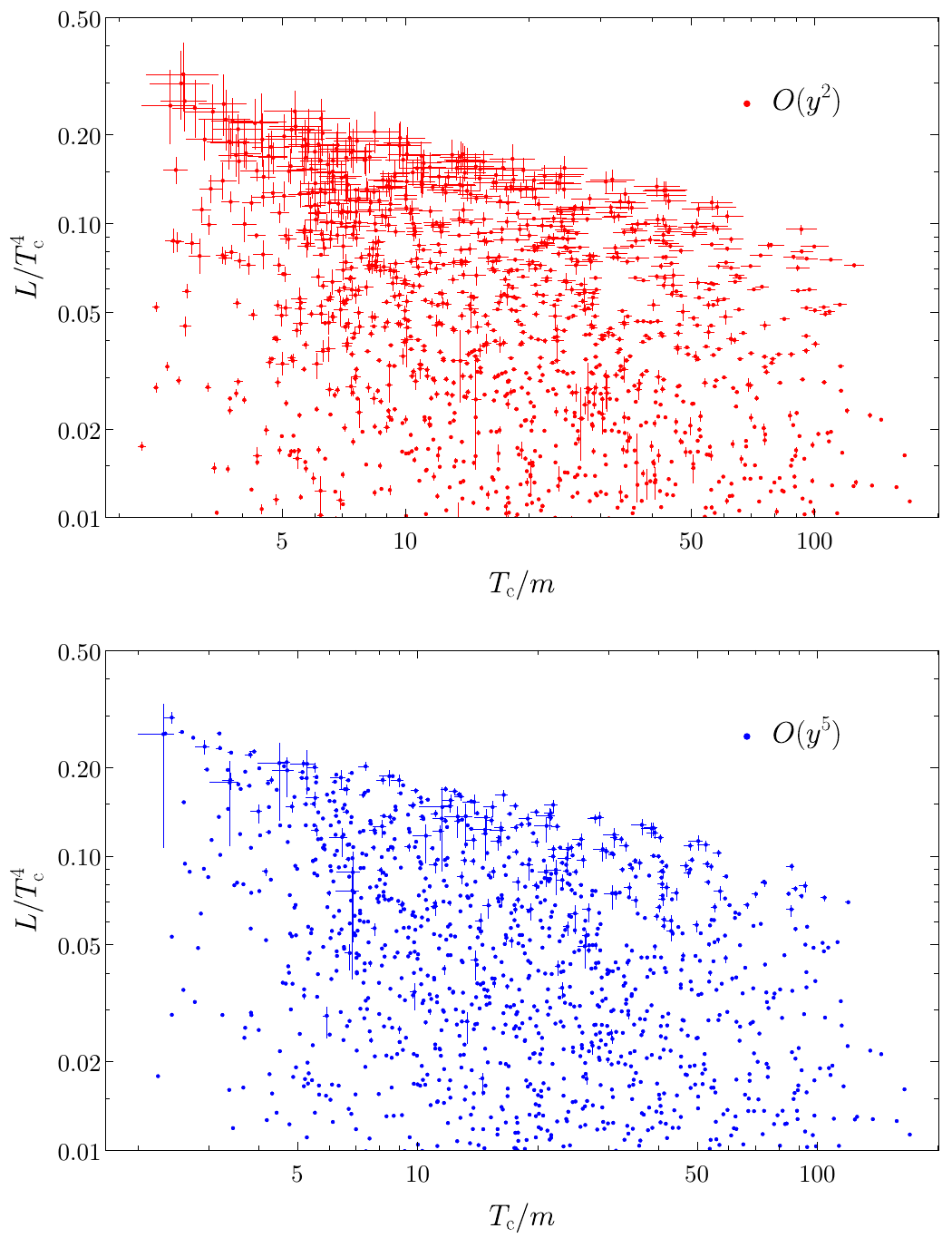}
\caption{$\Tc/m$ and $L/\Tc^4$ are plotted with their relative errors for theories that are \textit{consistent FOPTs} when evaluated with $f$ at both $\mathcal{O}({y^2T^4})$ and $\mathcal{O}({y^5T^4})$. There is a large reduction in relative error at higher order, especially for strong phase transitions, for our scan. There are also a number of points where the LO result underestimated the error.} 
\label{fig:TCLH_strong_LO_NLO_errors_compare}
\end{figure}

Figs.~\ref{fig:TCLH_LO_NLO_errors_compare} and \ref{fig:TCLH_strong_LO_NLO_errors_compare} show relative errors for theories that are \textit{consistent first-order phase transitions}, when evaluated at both $\mathcal{O}({y^2})$ and $\mathcal{O}({y^5})$ in the counting of eq.~\eqref{eq: y2 y5 count}. The uncertainties were determined by varying the renormalisation scale between $\mu=\pi T / 2$ and $\mu= 2 \pi T$.

Fig.~\ref{fig:TCLH_LO_NLO_errors_compare} shows that nearly all of the theory space benefits from reduced uncertainty at higher order. It is instructive to focus on the strongest first-order phase transitions, as only these are expected to yield gravitational wave signatures which are observable by next-generation detectors \cite{Gowling:2021gcy, Boileau:2022ter}. This can be seen in fig.~\ref{fig:TCLH_strong_LO_NLO_errors_compare}.

The errors are in general larger for the strongest transitions. For first-order phase transitions in our scan that are both strong (with $L/T^4 > 0.1$) and \textit{consistent} in both calculations, the higher order calculation reduces the average relative range in critical temperature from 18\% to 4\%, in latent heat from 19\% to 6\% and in $\drstar$ from 19\% to 7\%. If we considered even stronger phase transitions, with $L/T^4 > 0.2$, the higher order calculation reduces the average relative range in critical temperature from 28\% to 7\%, in latent heat from 40\% to 16\% and in $\drstar$ from 13\% to 5\%.

\begin{figure}[h]
 \captionsetup[subfigure]{margin={-0.9cm,0cm}}
\subfloat[\label{fig:benchmarks error tc}]{%
  \hspace*{0.5cm}\includegraphics[trim={2.9cm 7.4cm 5.3cm 2.5cm},clip,width=0.8\columnwidth, center]{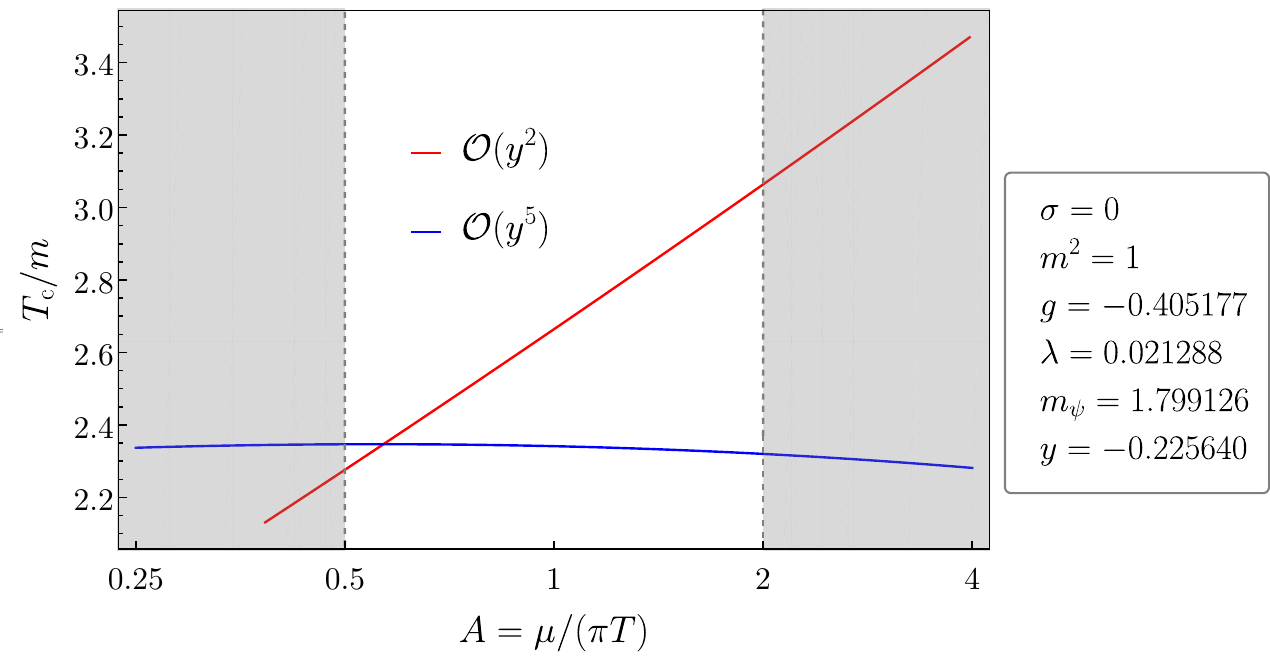}%
} 

\bigskip

\subfloat[\label{fig:benchmarks error L}]{%
  \hspace*{0.4cm}\includegraphics[trim={3cm 7.4cm 5.2cm 2.5cm},clip,width=0.8\columnwidth, center]{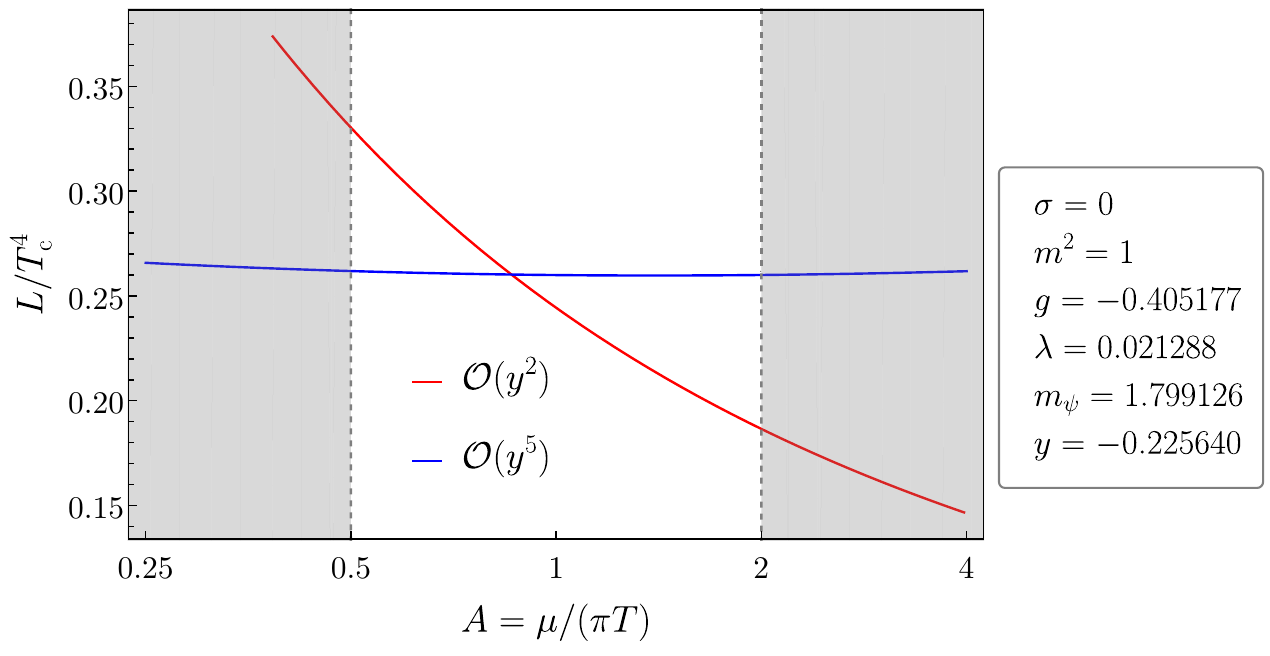}\hspace*{0.0cm}
} 
\caption{Uncertainties in $\Tc$ and $L/\Tc^4$ are significantly reduced when evaluating using the free energy density to $\mathcal{O}({y^5T^4})$\, here shown for a benchmark point with a strong transition.}
\label{fig:benchmark error reduction}
\end{figure}

Figure \ref{fig:benchmark error reduction} provides an example of a strong first order phase transition. The uncertainties in the critical temperature and the strength of the phase transition were significantly reduced at $\mathcal{O}({y^5})$ compared to the leading order $\mathcal{O}({y^2})$ evaluation. Note that below the scale choice $A \sim 0.4$, these quantities are not evaluated at $\mathcal{O}({y^2})$ as the parameter point is found to receive non-perturbative corrections to the critical temperature. Since this is outside of the range $1/2 \leq A \leq 2$, this parameter point is still considered a \textit{consistent first-order phase transition} at $\mathcal{O}({y^2})$.

As well as a reduction in uncertainty, it can be seen from figure \ref{fig:TCLH_strong_LO_NLO_errors_compare} that a small number of theories have larger estimated errors when evaluated for the higher order calculation. This reveals an underestimate of the uncertainty associated with the perturbative calculation when evaluated at leading order, and is another improvement offered by the calculation computing free energy density up to $\mathcal{O}({y^5T^4})$. Interestingly, the lower half of the $\mathcal{O}({y^2})$ plot resembles the $\mathcal{O}({y^5})$ plot, when the former is scaled up by a factor of $5\sim 10$, i.e.~an order of magnitude.

A ``conservative" approach was adopted here, where a first-order phase transitions is required to be consistently identified across a range of renormalisation scales. A similar story emerges if a more ``liberal" approach is taken, such as evaluating the nature of the theory only at a single renormalisation scale, for example setting $A=1$. Another option is to include a theory as a first-order phase transition if it is evaluated as such for any subset of the renormalisation scale range considered. Using each of these alternative approaches, applied to our scan, the calculation at $\mathcal{O}({y^5})$ evaluates additional strong phase transitions not identified at leading order using the same approach, combined with reduced relative error in the critical temperature and latent heat evaluated.

\section{Discussion}

Let us outline an answer to the central question of our study of cosmological phase transitions: how important are higher perturbative orders for a global parameter scan of a model? Previous studies have shown that higher orders are vital for making reliable predictions for cosmological phase transitions at individual benchmark parameter points in a model \cite{Kainulainen:2019kyp, Croon:2020cgk, Niemi:2020hto, Niemi:2021qvp, Gould:2021oba, Gould:2023ovu}. Yet it was unclear if these conclusions would extend to a global parameter scan, or whether higher perturbative orders would merely reshuffle parameter points.

We have tackled these questions using a Yukawa model as our guide. The simplicity of this model has allowed us to push to $\mathcal{O}({y^5 T^4})$ accuracy for the free-energy density. To do so, we have carried out high-temperature dimensional reduction, constructing a 3d effective field theory of only the lightest bosonic modes, and have included relevant infrared contributions up to three loops.

In short, higher orders reduce errors across the board, but are most important for describing the strongest phase transitions, where errors are largest. At leading order, the renormalisation scale dependence of the strongest phase transitions we have identified yield error estimates over 100\%, or change the character of the transition from first order through second order to crossover. Only at higher loops can the order of these strongest transitions, and their thermodynamic properties, be computed reliably.

This is expected to significantly impact predictions of the resulting gravitational wave signals, as these depend sensitively on the thermodynamics of the transition. Yet, additional nonequilibrium physics enters the computation of the gravitational wave spectrum, through the bubble nucleation rate and bubble wall speed. The calculation of higher orders for these quantities is in its infancy: for the nucleation rate the $\mathcal{O}(y^3)$ term has recently moved into reach \cite{Ekstedt:2021kyx, Gould:2021ccf, Ekstedt:2023sqc}, while for the bubble wall speed even a complete leading order calculation would extend the current state-of-the-art, which relies on a leading-log expansion \cite{Laurent:2022jrs}.

Future work could extend the model to include couplings to the Standard Model through the Higgs portal. Carrying through to explicitly evaluating the consequences for gravitational waves would then yield results directly applicable to the widely studied singlet fermionic dark matter candidate, extending the conclusions of previous works \cite{Fairbairn:2013uta, Li:2014wia, Beniwal:2018hyi} to higher orders. While the question remains as to how to compute higher orders for the nonequilibrium quantities which enter the gravitational wave spectrum, already some relevant information can be gleaned by upgrading the equilibrium calculation.

A natural question to ask next is: to what extent do we expect our conclusions to apply to other models? Crucially, strong first-order phase transitions are associated with large couplings and large errors also in scalar extensions of the Standard Model \cite{Caprini:2015zlo, Gould:2021oba, Niemi:2021qvp, Niemi:2020hto, Gould:2023ovu}, as well as in the Standard Model Effective Field Theory \cite{Croon:2020cgk}. Therefore, it seems plausible that the broad brushstrokes of our conclusions may apply to many other models, at least when the structure of the power counting relations is similar, that is, when the transitioning field is of the soft scale $m_\text{eff} \sim y T$. However, there may be qualitative differences for symmetry-breaking transitions, where the transitioning field instead lives at the supersoft scale $m_\text{eff} \sim y^{3/2} T / \pi$ \cite{Gould:2021ccf, Hirvonen:2022jba, Ekstedt:2022zro, Lofgren:2023sep}, or for very strong phase transitions where additional hierarchies of scale occur \cite{Kierkla:2022odc, Gould:2023ovu}.

\section*{Acknowledgments}
We would like to acknowledge insightful conversations with A.~Ekstedt, T.E.~Gonzalo, J.~Hirvonen, J.~L\"{o}fgren, 
B.~\'{S}wie\.{z}ewska and T.V.I.~Tenkanen.
We would also like to thank J.E.\ Camargo-Molina and the Swedish Collegium for Advanced Study for hosting the Thermal Field Theory and Early Universe Phenomenology at the Botanical Garden workshop, during which part of this work was completed.
O.G.~(ORCID ID 0000-0002-7815-3379) was supported by U.K.~Science and Technology Facilities Council (STFC) Consolidated Grant ST/T000732/1, a Research Leadership Award from the Leverhulme Trust, and a Royal Society Dorothy Hodgkin Fellowship.
C.X.~(ORCID ID 0000-0002-4886-9560) was supported by U.K.~STFC Consolidated Grant ST/S000666/1.

\appendix

\section{Power counting prescription} \label{appendix: power counting}

At high temperatures, thermal fluctuations (loops) qualitatively change the structure of the effective potential. At such temperatures, the vanilla loop expansion breaks down but one can still make progress by counting powers of couplings. To do so in a theory with multiple couplings, it is useful to fix power counting relations between the couplings.

To start with we assume that all three interaction terms are approximately equally perturbative. This assumption anyway underlies the usual loop expansion at zero temperature, and amounts to parametrically equating the loop expansion parameters for each interaction
\begin{equation} \label{eq:equally_perturbative}
\lambda \sim y^2 \sim \frac{g^2}{m^2}.
\end{equation}
The latter two terms are squared because each additional loop requires two cubic interactions, but only one quartic interaction, and follows from standard graph theoretic results \cite{Peskin:1995ev}.
The inverse powers of $m$ in the last term are necessary to cancel the dimensions of $g$, and arise from loop integrals of $\phi$. Note that we have not made explicit the factors of $1/(4\pi)$ arising from loop integrations.

In addition, we will assume that the leading effects of high-temperature fluctuations are of the same size as the tree-level scalar potential. For the scalar tadpole and mass, this amounts to
\begin{align}
    \frac{\sigma}{T^2} & \sim g \ , &
    \frac{m^2}{T^2} & \sim \lambda    \ .
\end{align}
This assumption is natural in the vicinity of a phase transition, as thermal fluctuations must balance against tree-level terms.

Finally, we must make an assumption about the fermion mass, $m_\psi$. For the fermion to have any effect on a phase transition, its mass cannot be large compared with the temperature. Otherwise its effects would be exponentially (Boltzmann) suppressed. For simplicity, we will power count the fermion mass the same as the scalar mass,
\begin{equation}
m_\psi \sim m \ .
\end{equation}
Our analysis will nevertheless include $m_\psi \ll m$ as a special case.

In summary, we adopt the following power counting prescription:
\begin{equation}
    \frac{\sigma}{T^3} \sim \frac{m^2}{T^2} \sim \frac{g}{T} \sim \lambda \sim y^2 \sim \frac{m_\psi^2}{T^2}  \ .
\end{equation}

A generic equilibrium observable $\theta$ has an expansion of the form $\theta/\theta_\text{LO} = 1 + c_1 y + c_2 y^2 + c_3 y^3 + \dots$, and in this work we calculate up to and including the $c_3 y^3$ term. For the effective potential and latent heat, which are $\mathcal{O}({y^2 T^4})$ at leading order, this means evaluation to $\mathcal{O}({y^5 T^4})$, or equivalently $\mathcal{O}({\lambda^{5/2} T^4})$, with errors of $\mathcal{O}({y^6 T^4})$. For the critical temperature, $\mathcal{O}({m y^{-1}})$ at leading order, the evaluation is to $\mathcal{O}({m y})$ as there is in general no contribution at $\mathcal{O}({my^2})$, with errors of $\mathcal{O}({my^3})$.

\section{Evaluating and matching correlation functions} \label{appendix: correlation functions}
We evaluated the connected, one-particle irreducible correlation functions, denoted by $\Gamma^{(k)}$ in the 4d theory and $\Gamma_3^{(k)}$ in the 3d effective theory. These observables were then be used to derive matching relations between the theories to find expressions for the effective parameters.

In constructing the EFT~\cite{Manohar:2018aog, Burgess:2020tbq}, we are aiming to capture the averaged effect of the hard scale modes ($\sim \pi T$) on the soft scale modes ($\sim y T$). As such, in the computation of correlation functions we are free to cut off the soft modes in any way, as long as we do so on both sides of the matching relations. This observation underlies the \textit{strict perturbation theory} approach of Braaten and Nieto \cite{Braaten:1995cm}, which we follow. Essentially one expands all loop integrals in powers of the soft scale, and then uses dimensional regularisation to cut off any infrared divergences. In this context, matching correlation functions is equivalent to directly integrating out the hard modes \cite{Burgess:2020tbq, Hirvonen:2022jba}.

\begin{figure}[ht]
    \centering
    \begin{subfigure}[b]{0.16\textwidth}
    \centering
    \includegraphics[width=\textwidth]{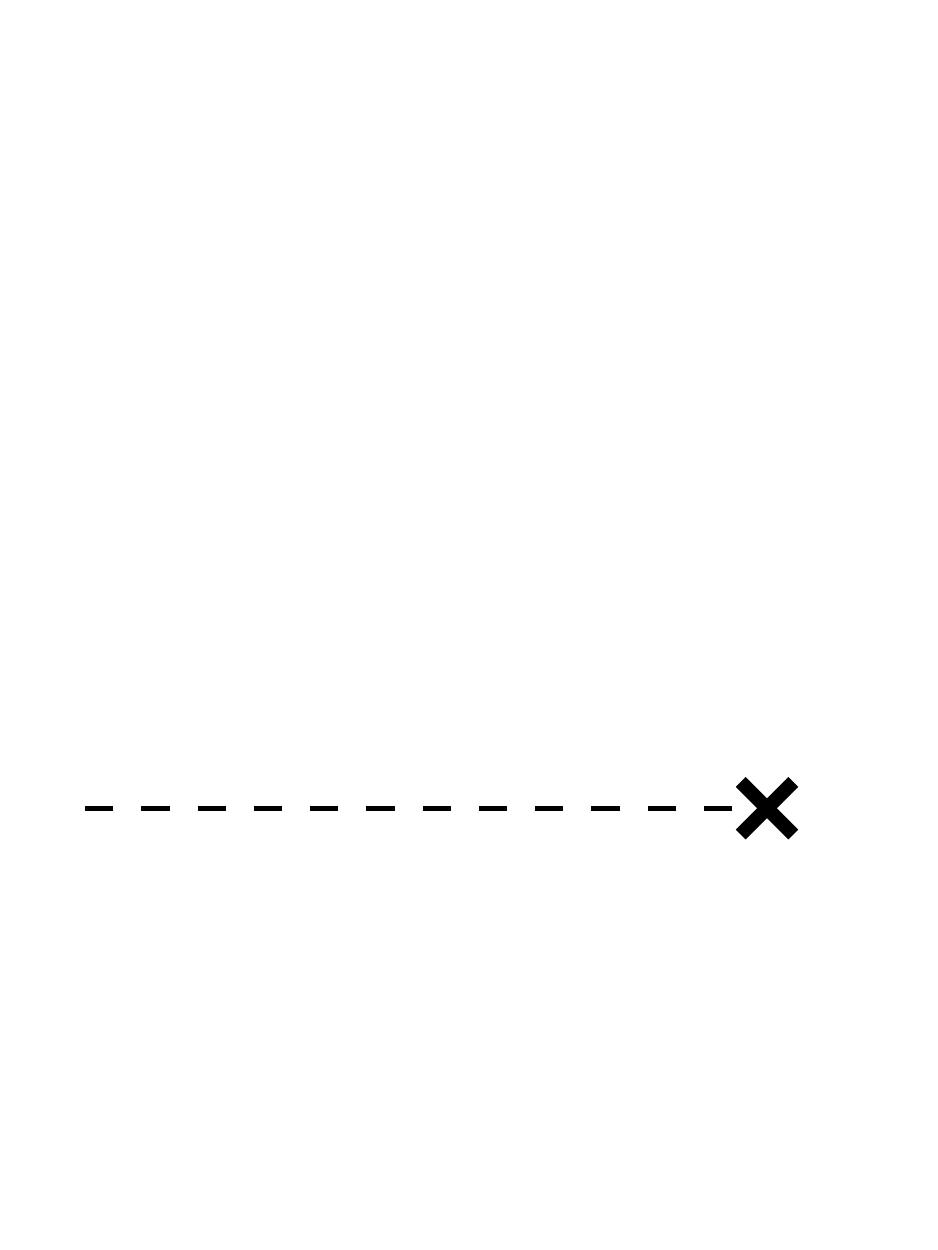}
    \caption{}
    \end{subfigure}
    \hfill
    \begin{subfigure}[b]{0.16\textwidth}
    \centering
    \includegraphics[width=\textwidth]{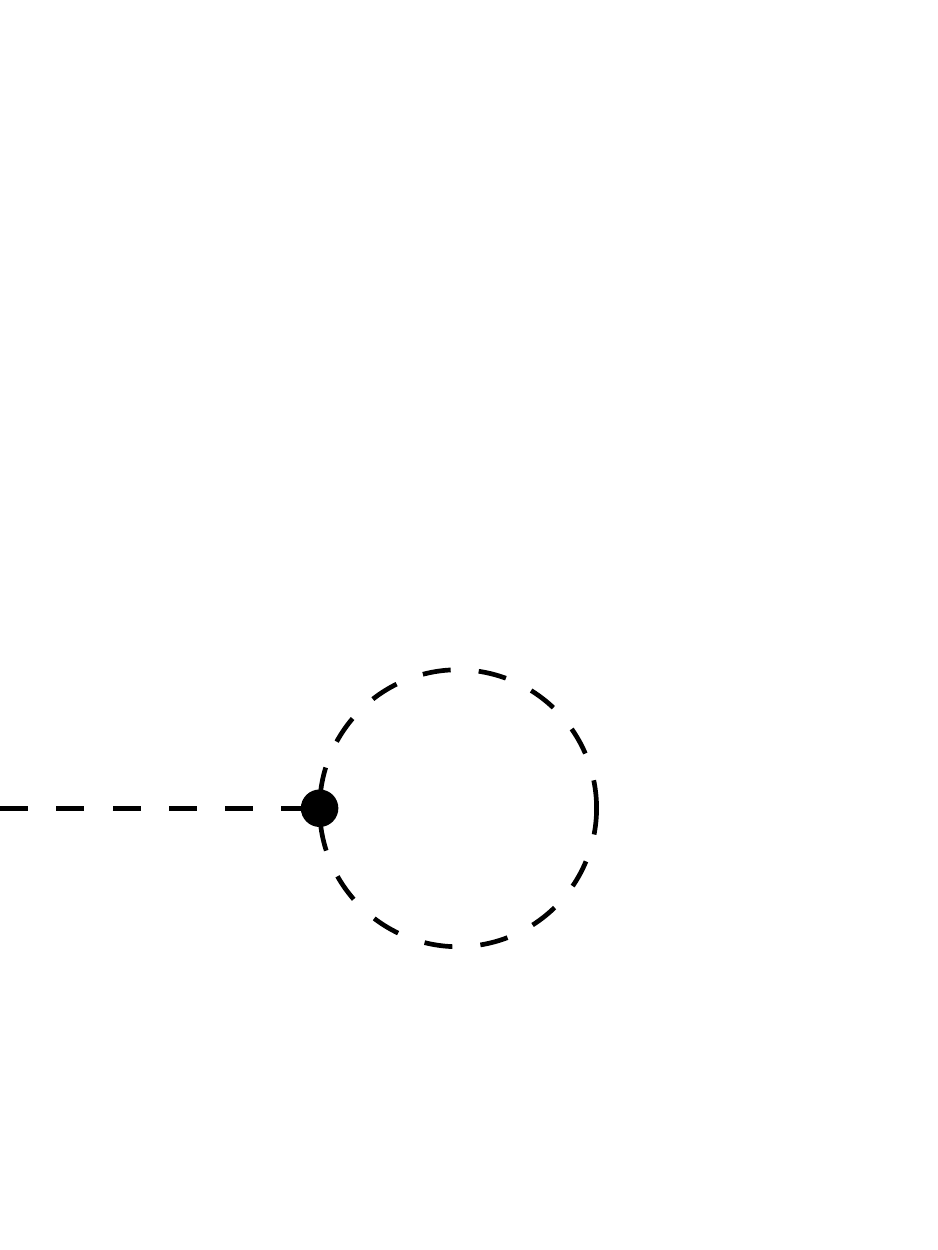}
    \caption{}
    \end{subfigure}
    \hfill
    \begin{subfigure}[b]{0.16\textwidth}
    \centering
    \includegraphics[width=\textwidth]{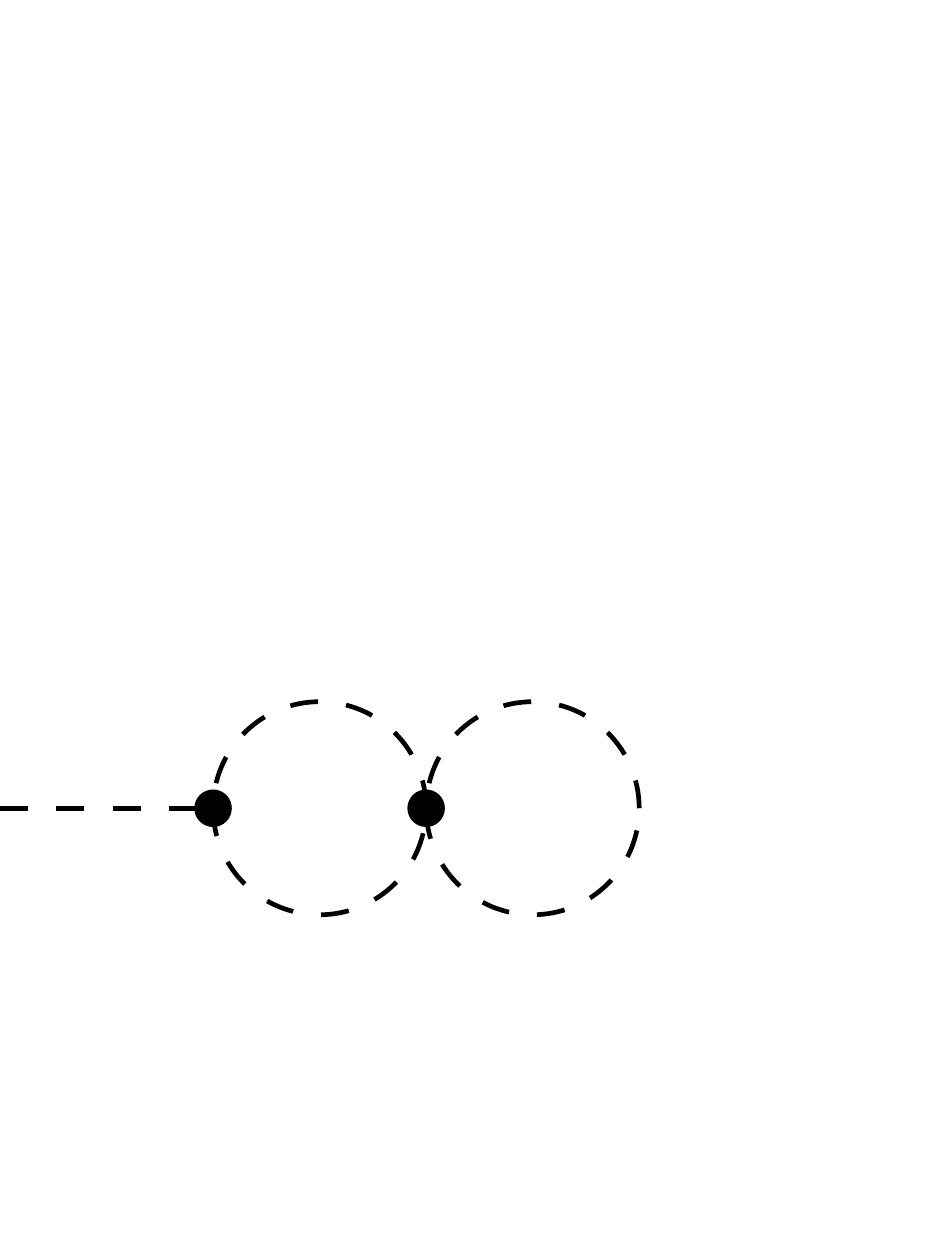}
    \caption{}
    \end{subfigure}
    \hfill
    \begin{subfigure}[b]{0.16\textwidth}
    \centering
    \includegraphics[width=\textwidth]{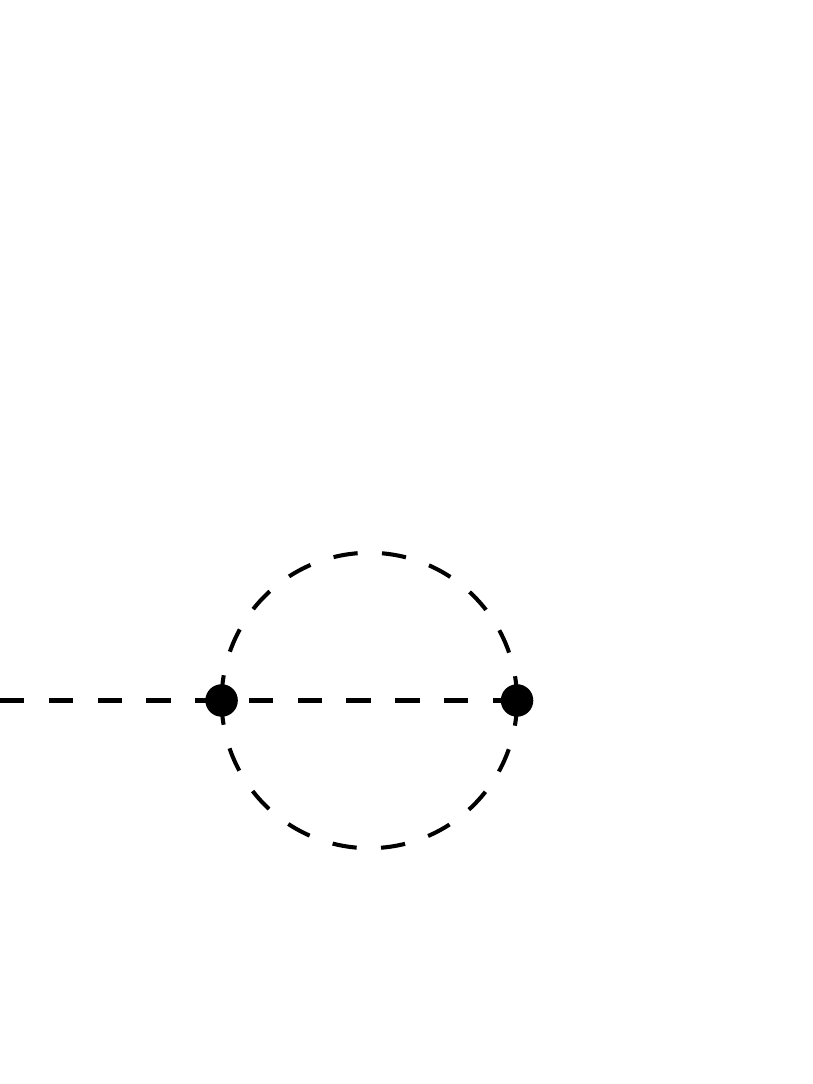}
    \caption{}
    \end{subfigure}
    \hfill
    \begin{subfigure}[b]{0.16\textwidth}
    \centering
    \includegraphics[width=\textwidth]{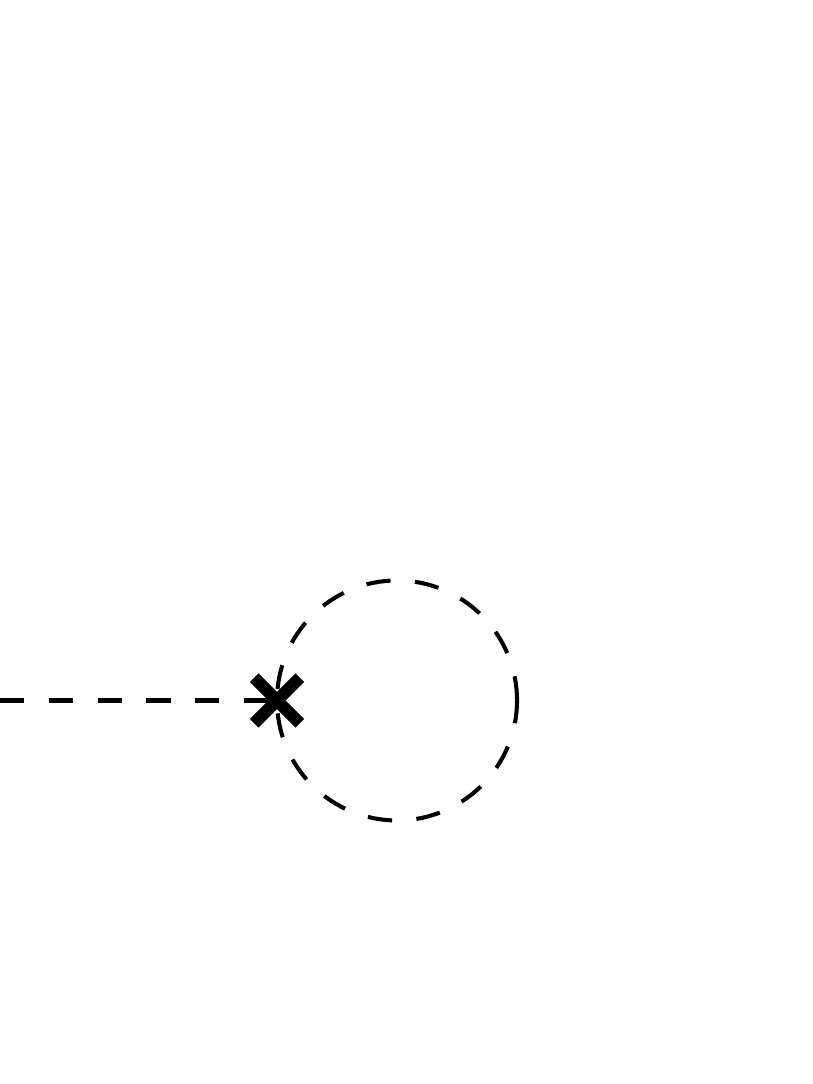}
    \caption{}
    \end{subfigure}
    \hfill
    \begin{subfigure}[b]{0.16\textwidth}
    \centering
    \includegraphics[width=\textwidth]{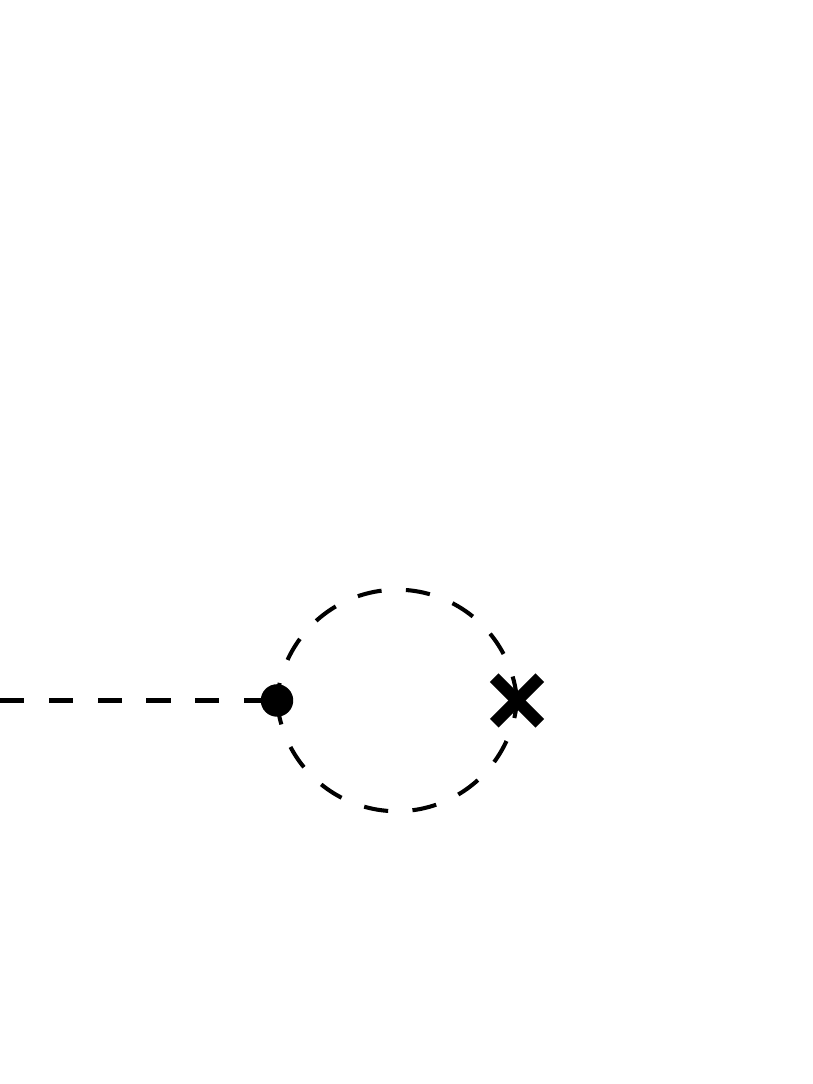}
    \caption{}
    \end{subfigure}
    
    \begin{subfigure}[b]{0.16\textwidth}
    \centering
    \includegraphics[width=\textwidth]{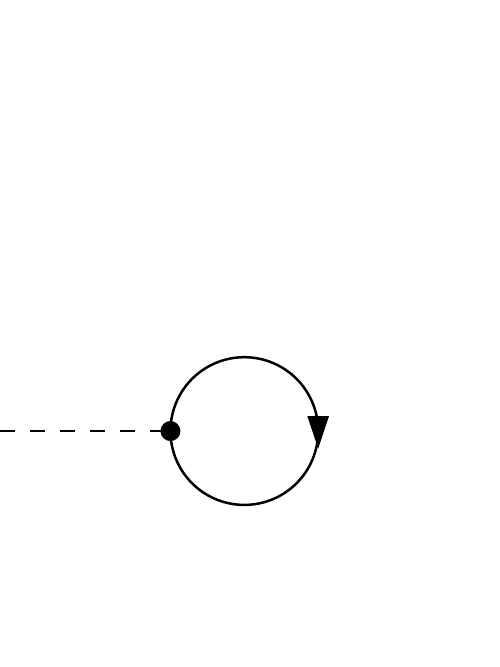}
    \caption{}
    \end{subfigure}
    \hfill
    \begin{subfigure}[b]{0.16\textwidth}
    \centering
    \includegraphics[width=\textwidth]{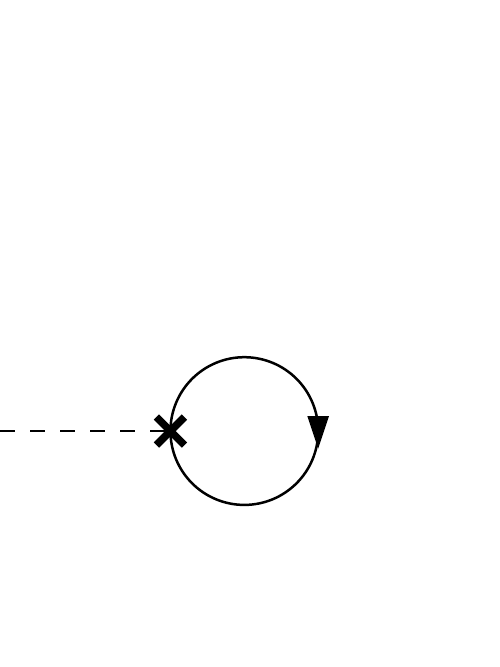}
    \caption{}
    \end{subfigure}
    \hfill
    \begin{subfigure}[b]{0.16\textwidth}
    \centering
    \includegraphics[width=\textwidth]{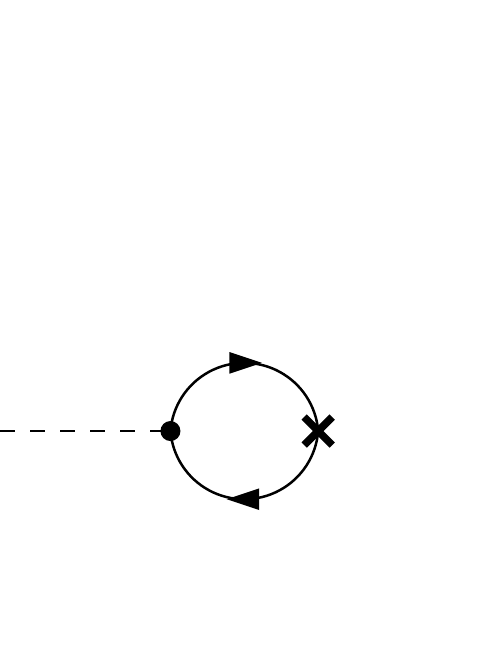}
    \caption{}
    \end{subfigure}
    \hfill
    \begin{subfigure}[b]{0.16\textwidth}
    \centering
    \includegraphics[width=\textwidth]{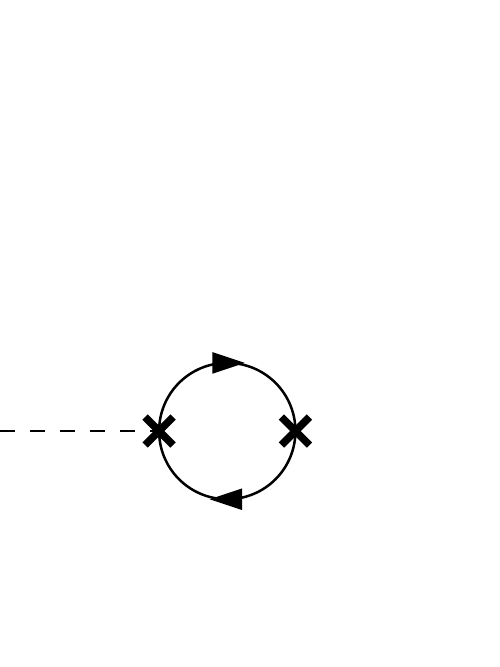}
    \caption{}
    \end{subfigure}
    \hfill
    \begin{subfigure}[b]{0.16\textwidth}
    \centering
    \includegraphics[width=\textwidth]{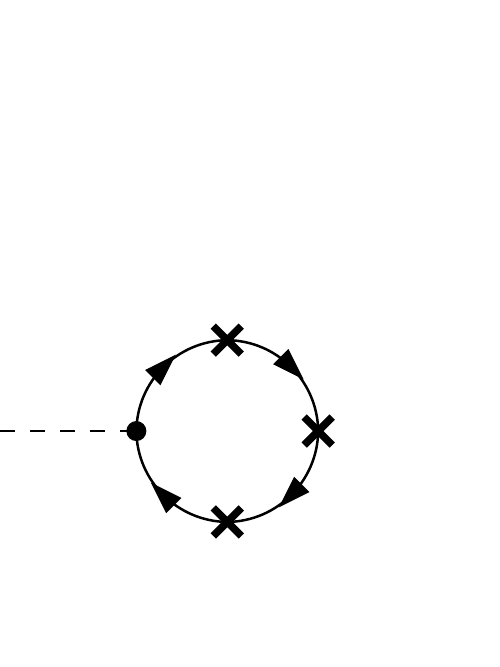}
    \caption{}
    \end{subfigure}
    \hfill
    \begin{subfigure}[b]{0.16\textwidth}
    \centering
    \includegraphics[width=\textwidth]{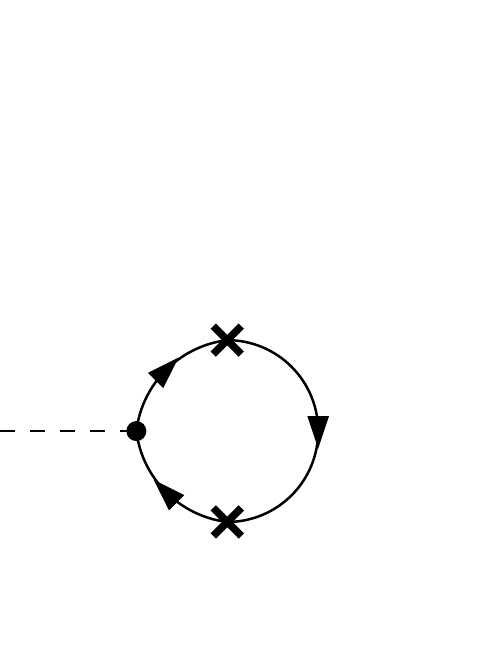}
    \caption{}
    \end{subfigure}
        
    \begin{subfigure}[b]{0.16\textwidth}
    \centering
    \includegraphics[width=\textwidth]{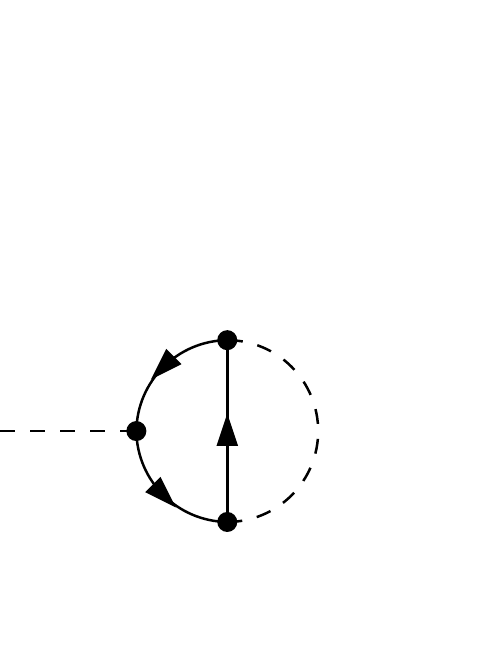}
    \caption{}
    \end{subfigure}
    \hspace{0.0\textwidth}
    \begin{subfigure}[b]{0.16\textwidth}
    \centering
    \includegraphics[width=\textwidth]{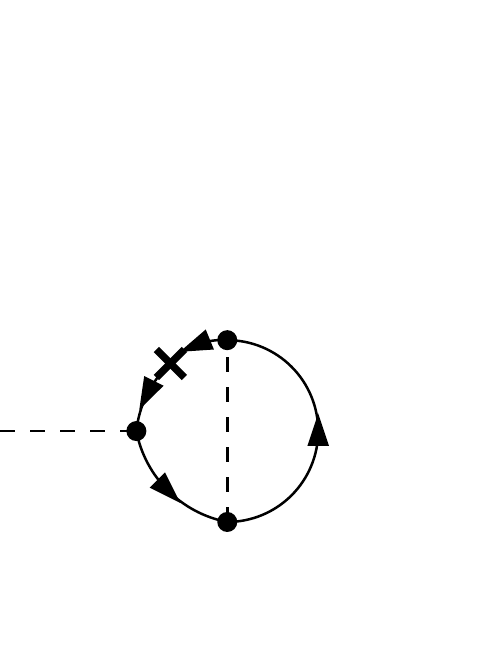}
    \caption{}
    \end{subfigure}
    \hspace{0.0\textwidth}
    \begin{subfigure}[b]{0.16\textwidth}
    \centering
    \includegraphics[width=\textwidth]{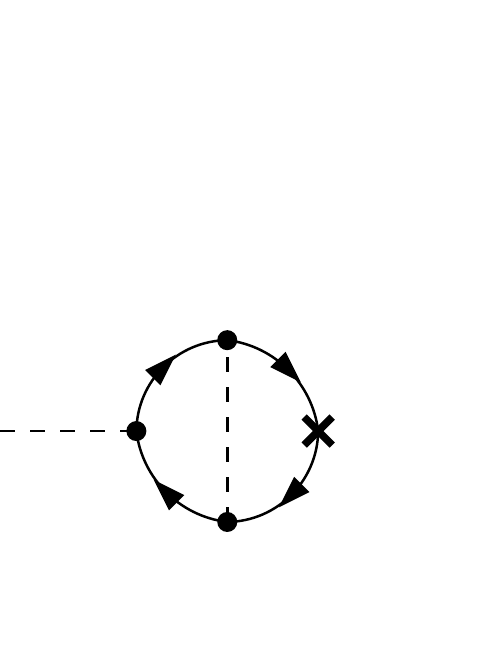}
    \caption{}
    \end{subfigure}
    \hspace{0.0\textwidth}
    \begin{subfigure}[b]{0.16\textwidth}
    \centering
    \includegraphics[width=\textwidth]{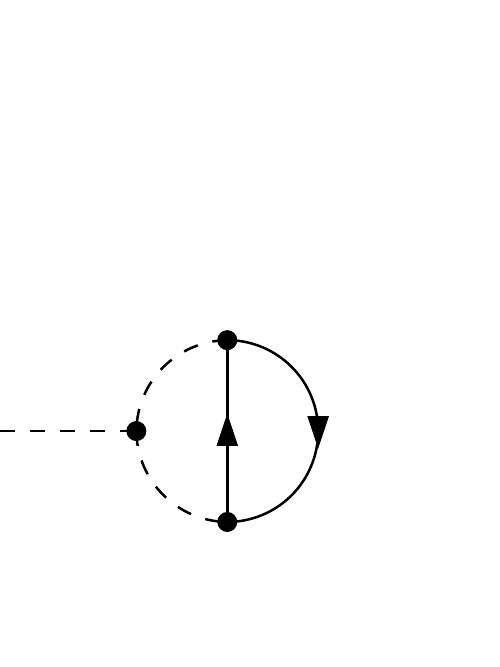}
    \caption{}
    \end{subfigure}
    
    \caption{The Feynman diagrams contributing to the one-point correlation function, $\Gamma^{(1)}(\mathbf{0})$ up to $\mathcal{O}(y^4 T^{3})$, with power counting set out in eq.~\eqref{eq: power counting prescription}.}
    \label{fig:gamma1}
\end{figure}

\begin{figure}[ht]
    \centering
    \begin{subfigure}[b]{0.16\textwidth}
    \centering
    \includegraphics[width=\textwidth]{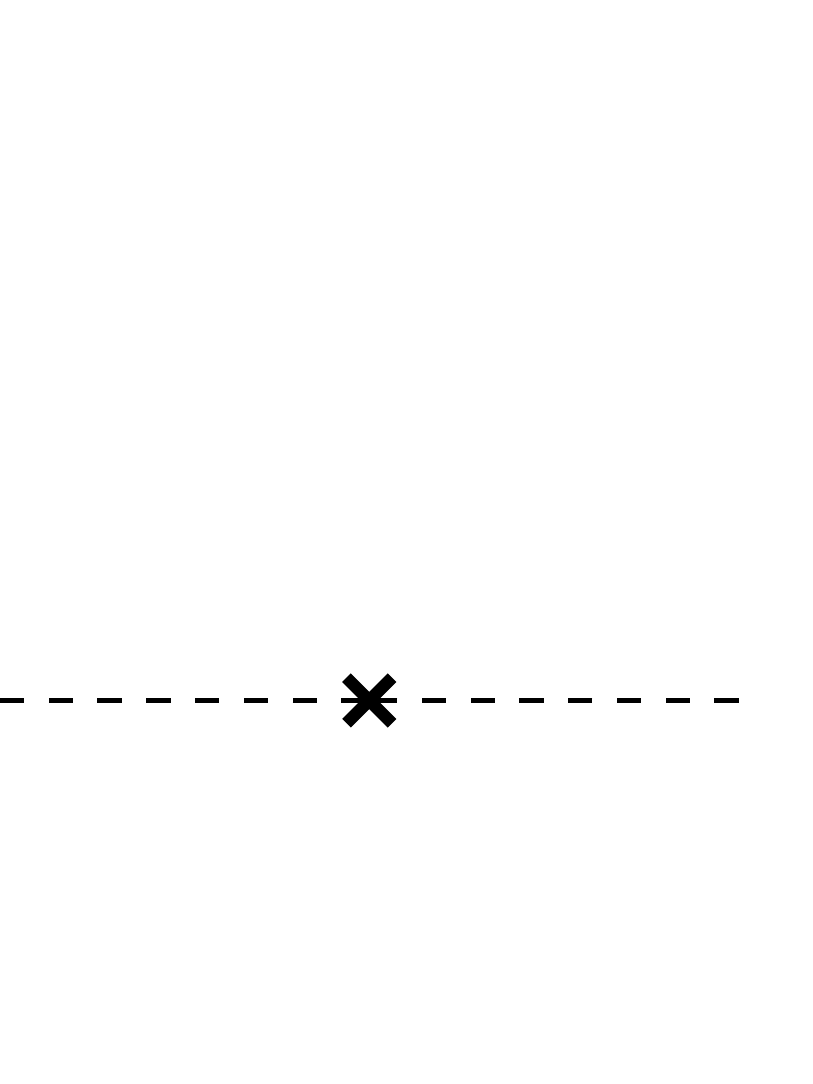}
    \caption{}
    \end{subfigure}
    \hfill
    \begin{subfigure}[b]{0.16\textwidth}
    \centering
    \includegraphics[width=\textwidth]{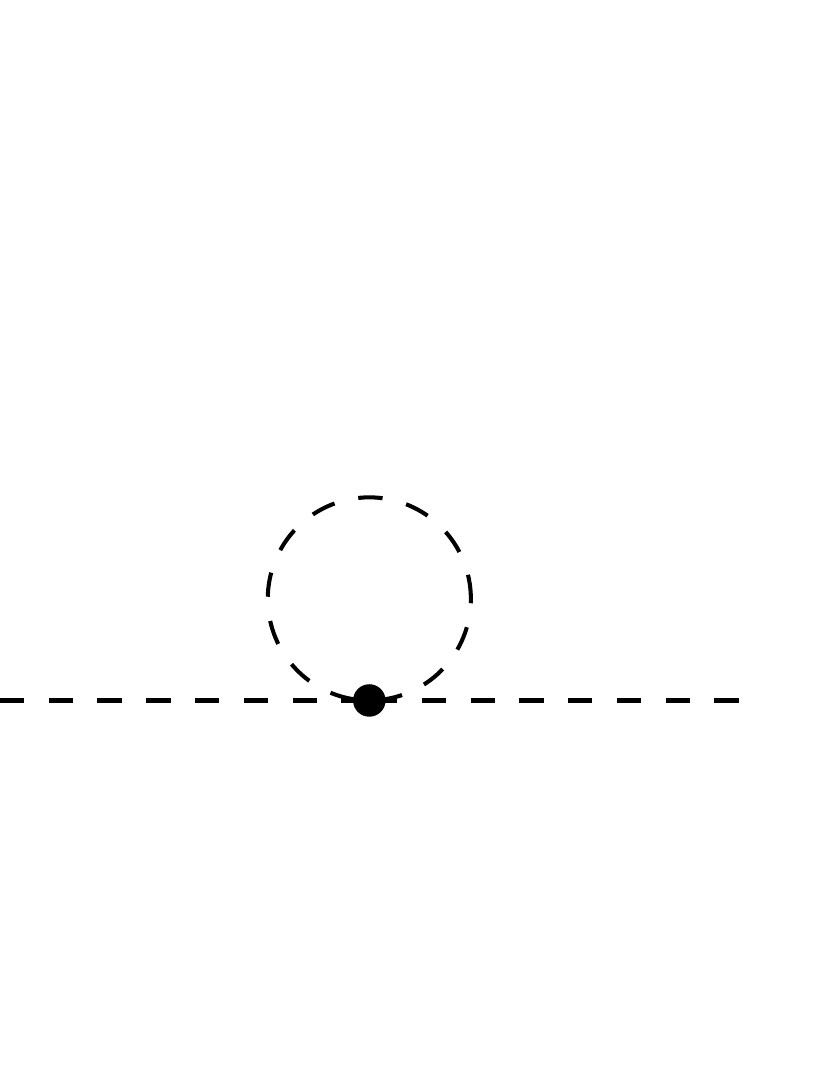}
    \caption{}
    \end{subfigure}
    \hfill
    \begin{subfigure}[b]{0.16\textwidth}
    \centering
    \includegraphics[width=\textwidth]{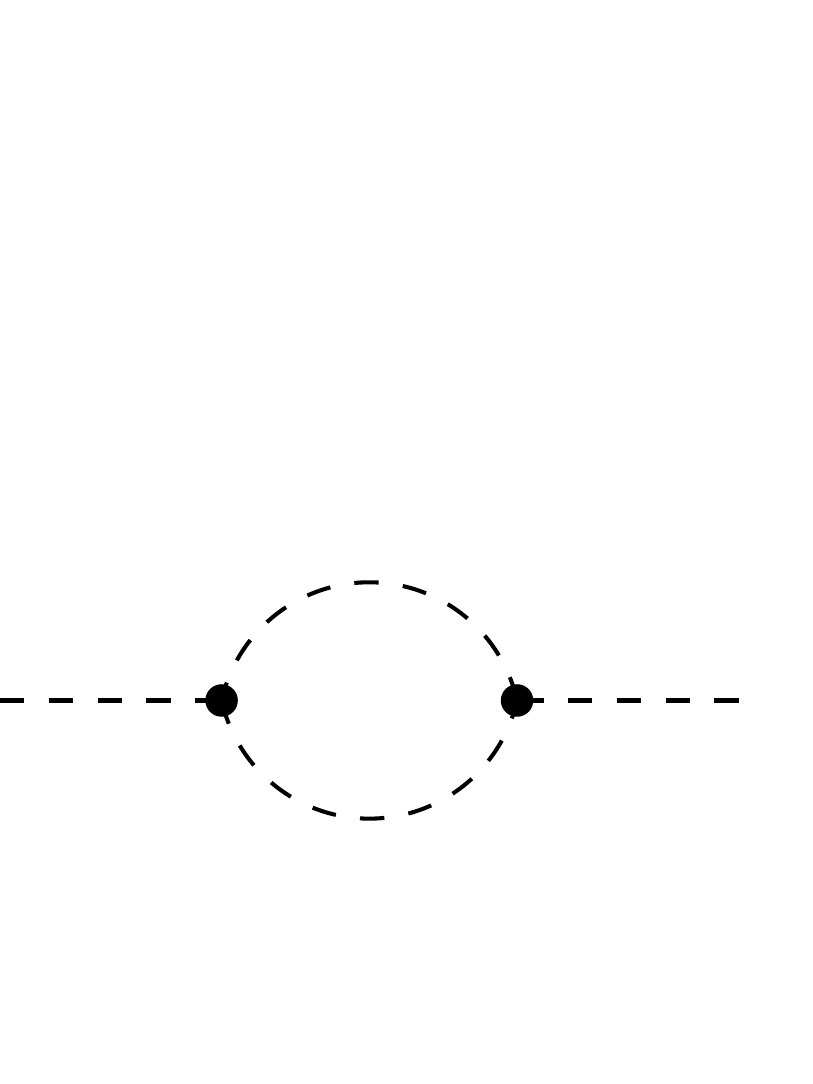}
    \caption{}
    \end{subfigure}
    \hfill
    \begin{subfigure}[b]{0.16\textwidth}
    \centering
    \includegraphics[width=\textwidth]{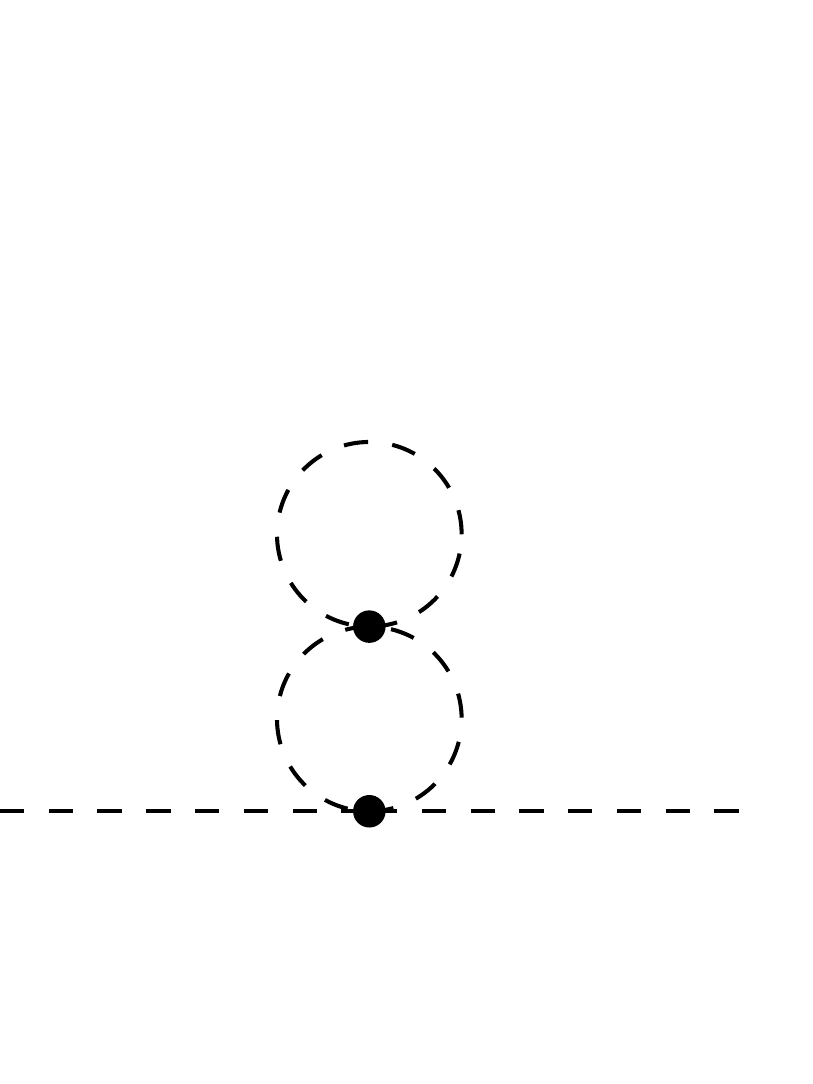}
    \caption{}
    \end{subfigure}
    \hfill
    \begin{subfigure}[b]{0.16\textwidth}
    \centering
    \includegraphics[width=\textwidth]{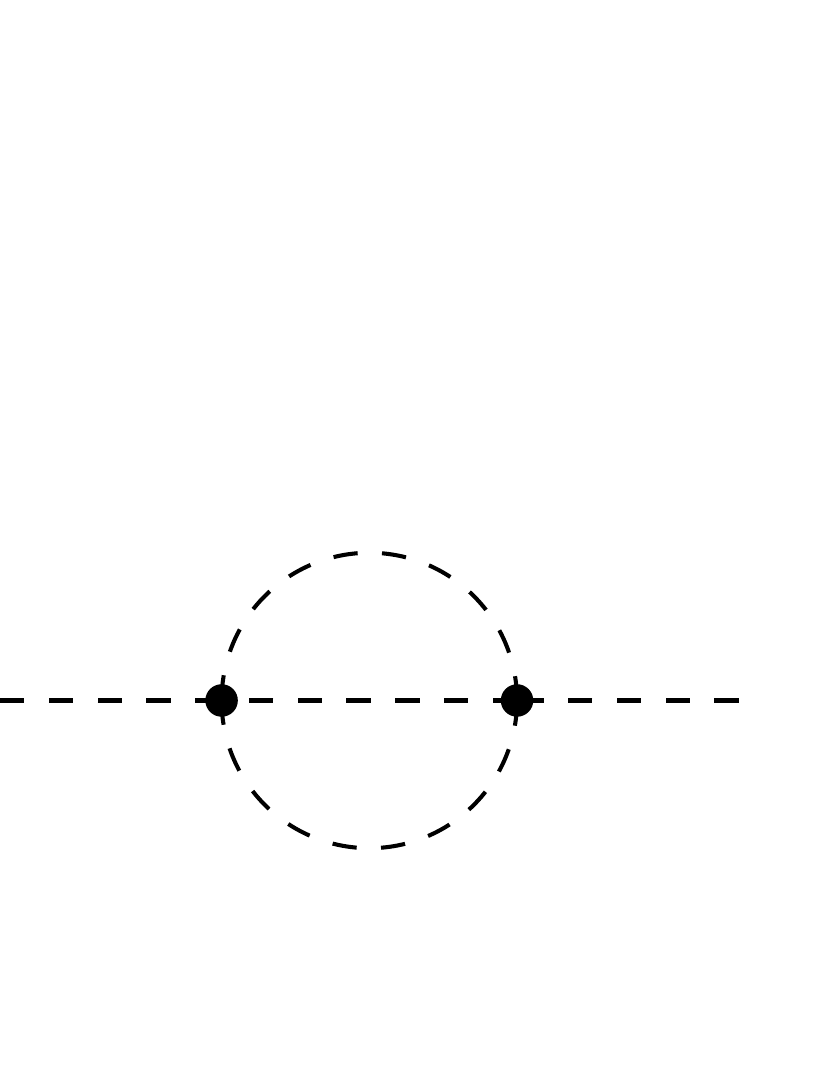}
    \caption{}
    \end{subfigure}
    \hfill
    \begin{subfigure}[b]{0.16\textwidth}
    \centering
    \includegraphics[width=\textwidth]{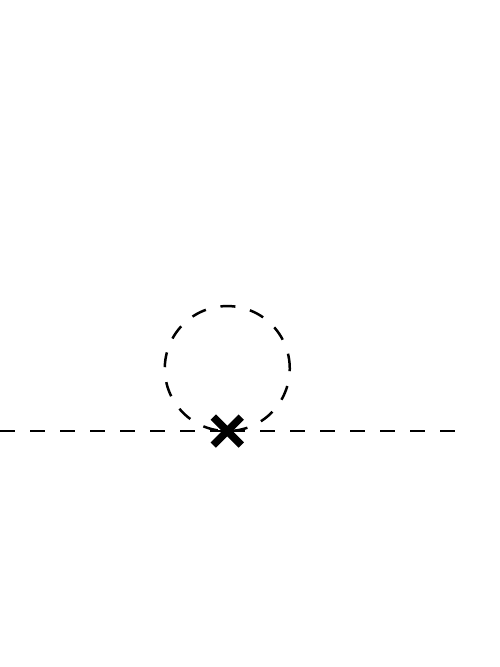}
    \caption{}
    \end{subfigure}

    \begin{subfigure}[b]{0.16\textwidth}
    \centering
    \includegraphics[width=\textwidth]{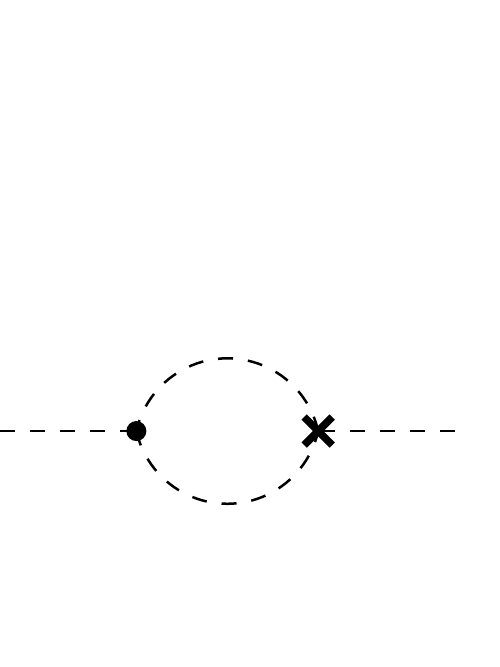}
    \caption{}
    \end{subfigure}
    \hfill
    \begin{subfigure}[b]{0.16\textwidth}
    \centering
    \includegraphics[width=\textwidth]{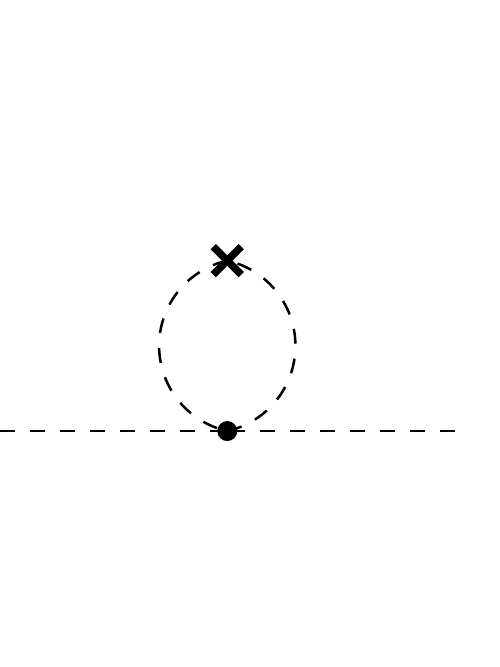}
    \caption{}
    \end{subfigure}
    \hfill
    \begin{subfigure}[b]{0.16\textwidth}
    \centering
    \includegraphics[width=\textwidth]{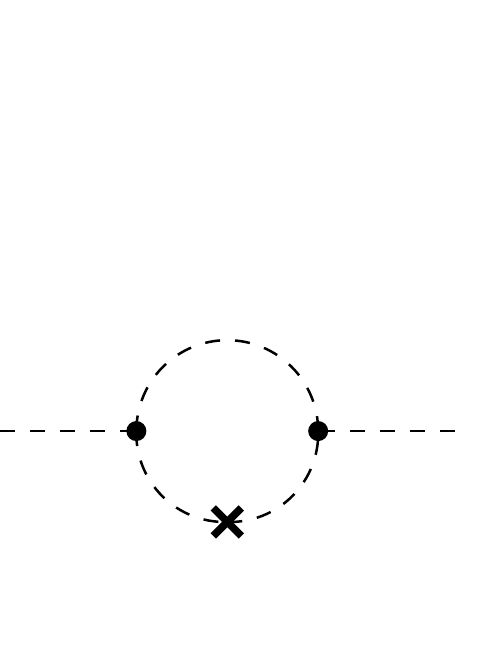}
    \caption{}
    \end{subfigure}
    \hfill
    \begin{subfigure}[b]{0.16\textwidth}
    \centering
    \includegraphics[width=\textwidth]{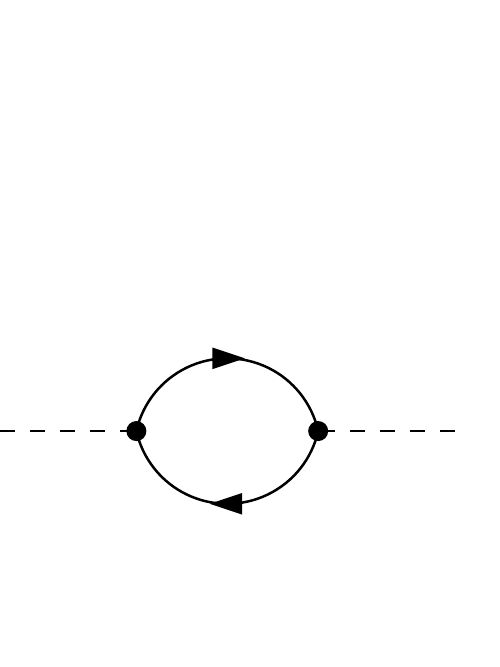}
    \caption{}
    \end{subfigure}
    \hfill
    \begin{subfigure}[b]{0.16\textwidth}
    \centering
    \includegraphics[width=\textwidth]{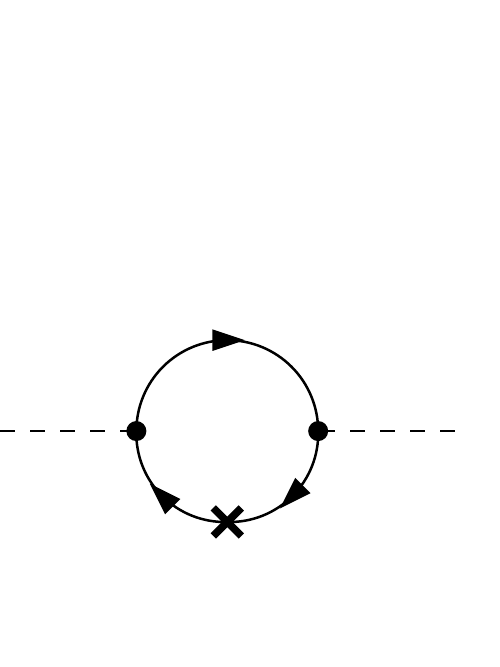}
    \caption{}
    \end{subfigure}
    \hfill
    \begin{subfigure}[b]{0.16\textwidth}
    \centering
    \includegraphics[width=\textwidth]{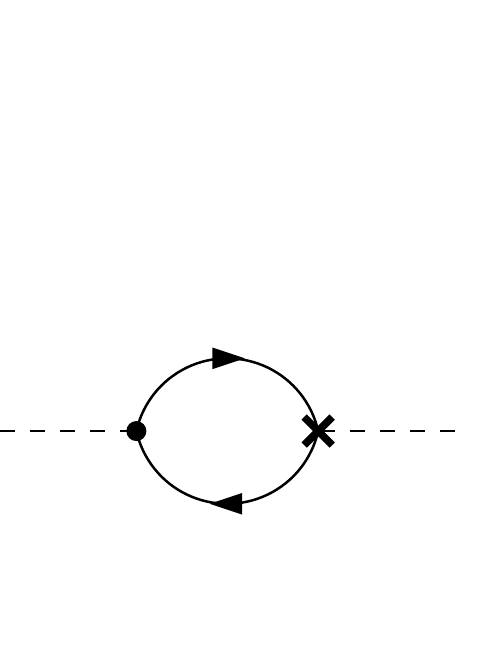}
    \caption{}
    \end{subfigure}
    
    \begin{subfigure}[b]{0.16\textwidth}
    \centering
    \includegraphics[width=\textwidth]{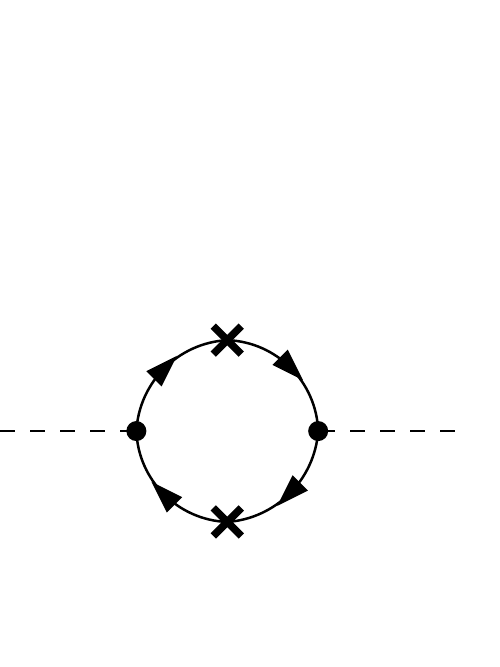}
    \caption{}
    \end{subfigure}
    \begin{subfigure}[b]{0.16\textwidth}
    \centering
    \includegraphics[width=\textwidth]{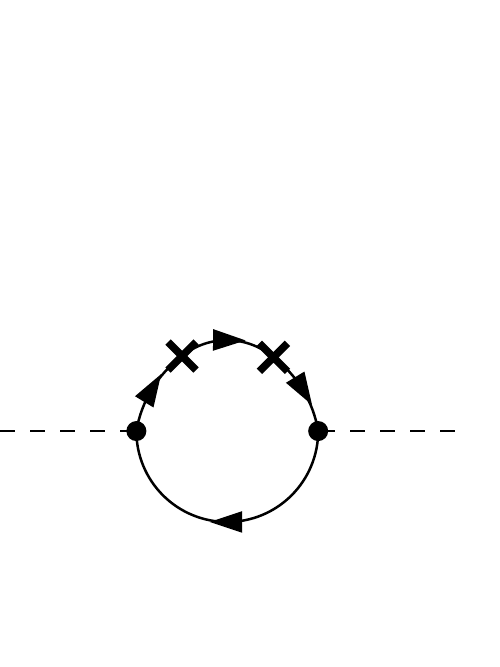}
    \caption{}
    \end{subfigure}
    \hfill
    \begin{subfigure}[b]{0.16\textwidth}
    \centering
    \includegraphics[width=\textwidth]{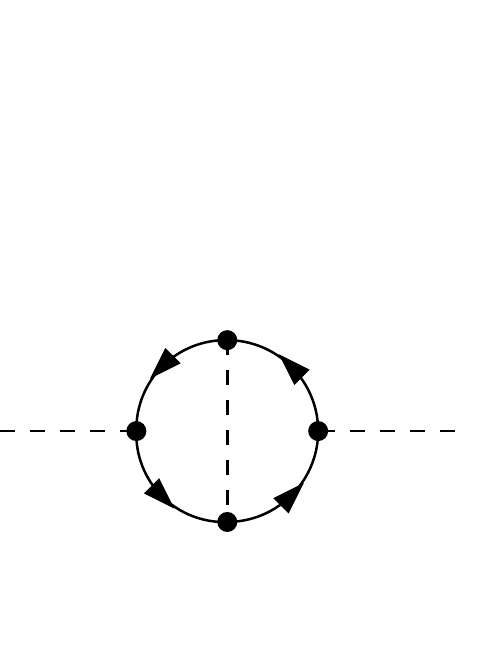}
    \caption{}
    \end{subfigure}
    \hfill
    \begin{subfigure}[b]{0.16\textwidth}
    \centering
    \includegraphics[width=\textwidth]{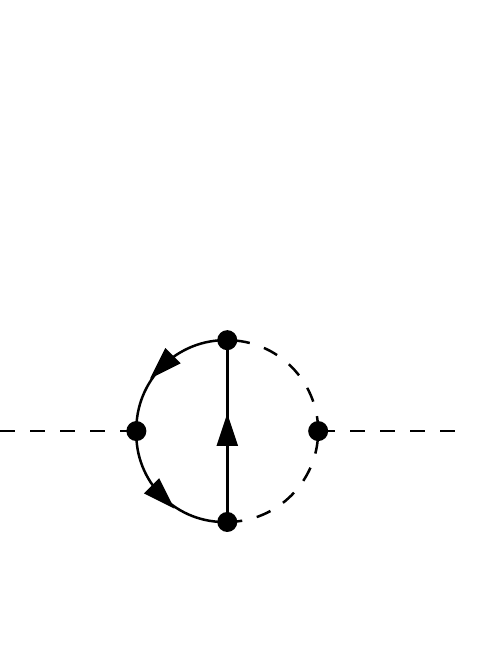}
    \caption{}
    \end{subfigure}
    \hfill
    \begin{subfigure}[b]{0.16\textwidth}
    \centering
    \includegraphics[width=\textwidth]{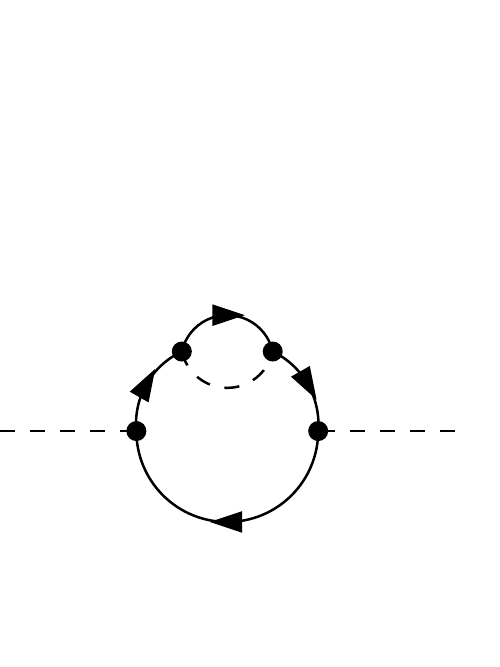}
    \caption{}
    \end{subfigure}
    \hfill
    \begin{subfigure}[b]{0.16\textwidth}
    \centering
    \includegraphics[width=\textwidth]{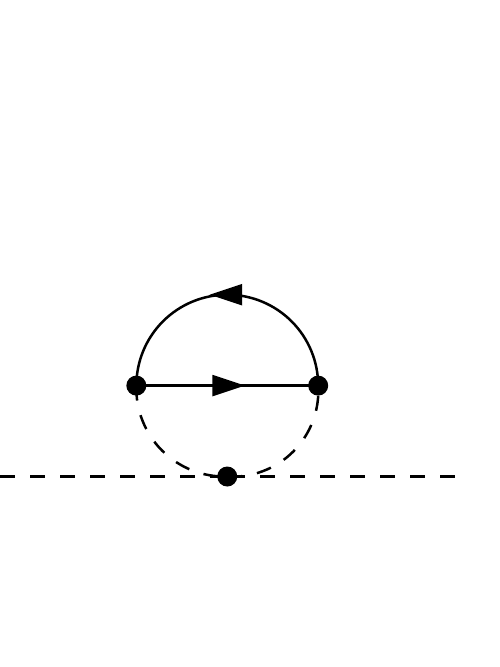}
    \caption{}
    \end{subfigure}  

    \caption{The Feynman diagrams contributing to the two-point correlation function, $\Gamma^{(2)}(\mathbf{p,-p})$ up to $\mathcal{O}(y^4 T^{2})$.}
    \label{fig:gamma2}
\end{figure}

\begin{figure}[ht]
    \centering
    \begin{subfigure}[b]{0.16\textwidth}
    \centering
    \includegraphics[width=\textwidth]{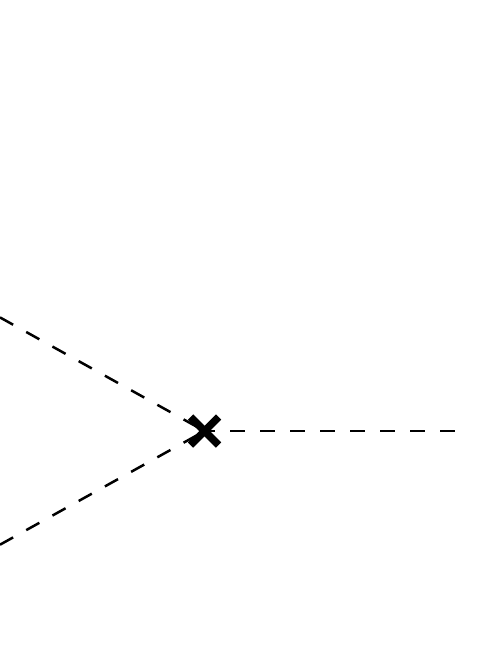}
    \caption{}
    \end{subfigure}
    \hspace{0.01\textwidth}
    \begin{subfigure}[b]{0.16\textwidth}
    \centering
    \includegraphics[width=\textwidth]{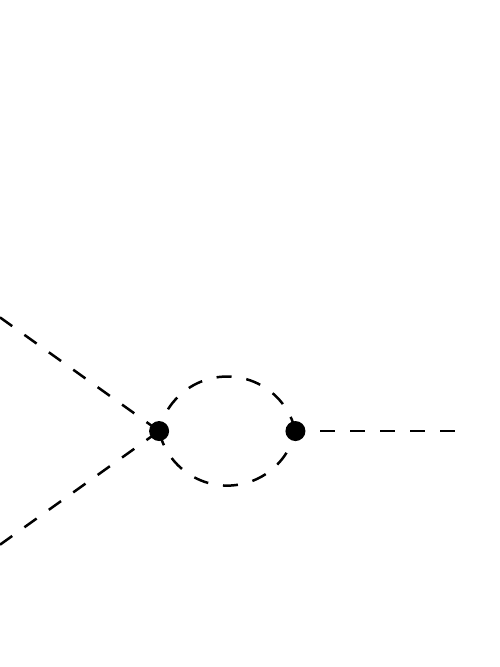}
    \caption{}
    \end{subfigure}
    \hspace{0.01\textwidth}
    \begin{subfigure}[b]{0.16\textwidth}
    \centering
    \includegraphics[width=\textwidth]{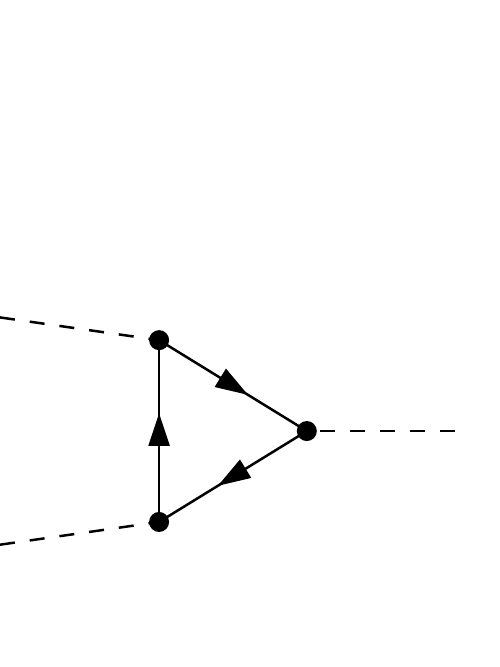}
    \caption{}
    \end{subfigure}
    \hspace{0.01\textwidth}
    \begin{subfigure}[b]{0.16\textwidth}
    \centering
    \includegraphics[width=\textwidth]{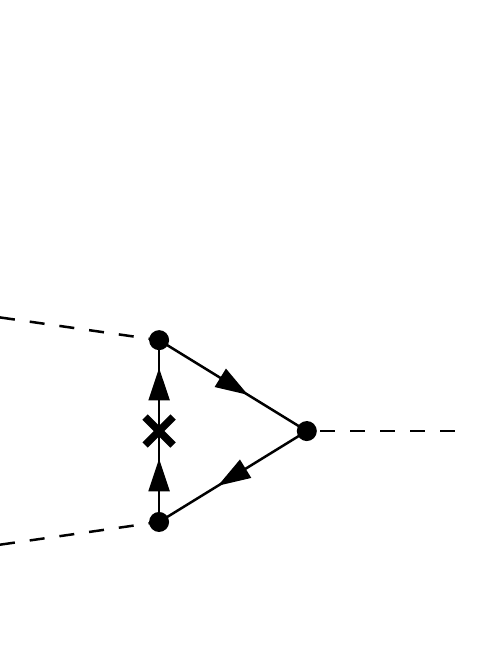}
    \caption{}
    \end{subfigure}
    \caption{The Feynman diagrams contributing to the three-point correlation function, $\Gamma^{(3)}(\mathbf{p,q,-p,-q})$ up to $\mathcal{O}(y^4 T)$.}
    \label{fig:gamma3}
\end{figure}

\begin{figure}[ht]
\centering
\begin{subfigure}[b]{0.16\textwidth}
\centering
\includegraphics[width=\textwidth]{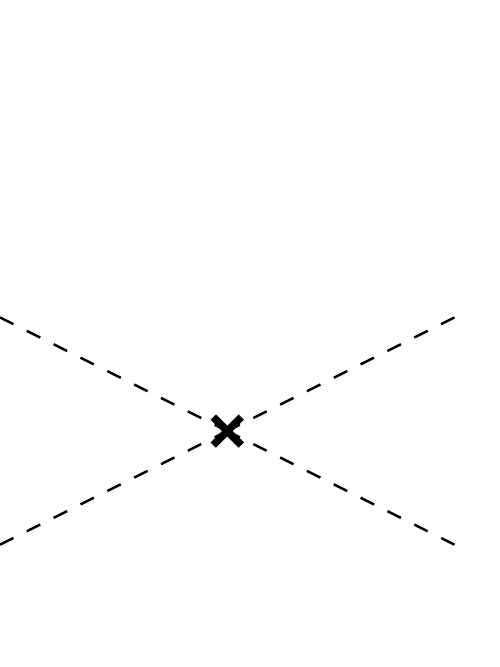}
\caption{}
\end{subfigure}
\hspace{0.01\textwidth}
\begin{subfigure}[b]{0.16\textwidth}
\centering
\includegraphics[width=\textwidth]{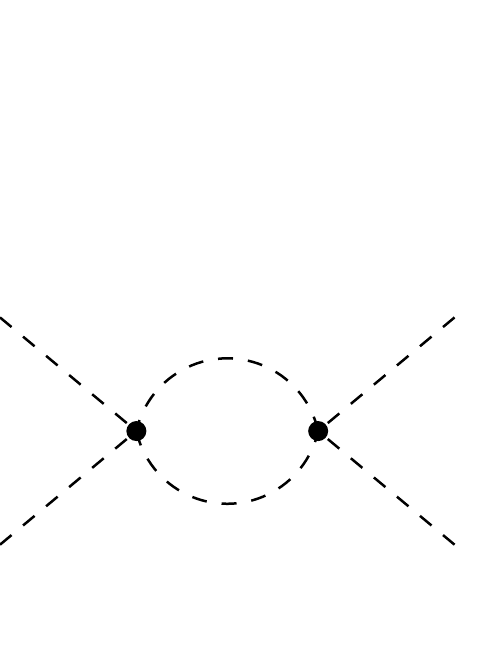}
\caption{}
\end{subfigure}
\hspace{0.01\textwidth}
\begin{subfigure}[b]{0.16\textwidth}
\centering
\includegraphics[width=\textwidth]{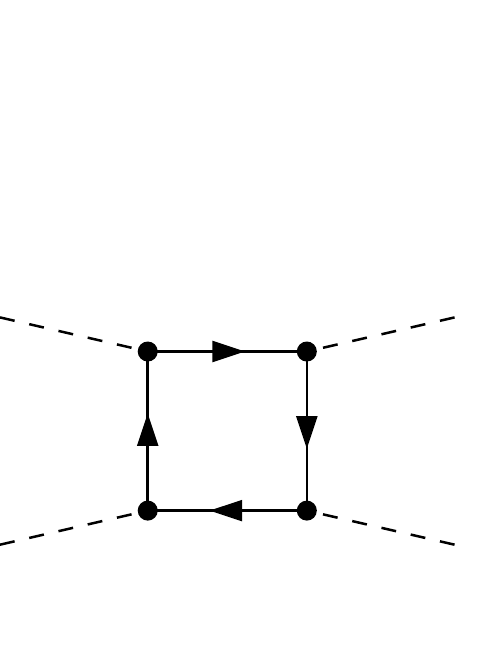}
\caption{}
\end{subfigure}

    \caption{The Feynman diagrams contributing to the four-point correlation function, $\Gamma^{(4)}(\mathbf{p,q,r,-p,-q,-r})$ up to $\mathcal{O}(y^4)$.}
    \label{fig:gamma4}
\end{figure}

The one, two, three and four-point correlation functions are given by the sum of diagrams in figs.~\ref{fig:gamma1}, \ref{fig:gamma2}, \ref{fig:gamma3} and \ref{fig:gamma4} respectively. These were generated using FeynArts~\cite{Kublbeck:1990xc}. Only diagrams and terms up to order $\mathcal{O}(y^5)$ in the power counting prescription eq.~\eqref{eq: power counting prescription} were included in the calculations. Tadpole and mass terms are treated as interactions, according to strict perturbation theory, due to the hierarchy of energy scales between their coefficients and $\pi T$ which characterises the free Lagrangian. The results of the loop sum integrals can be found in the literature and are listed in appendix \ref{appendix: loop integrals}. 

The evaluation of the correlation functions are given in equations \eqref{eq:gamma1_integrals} to \eqref{eq:gamma4_integrated}. The Euler-Mascheroni constant is denoted by $\gamma_E$, and $A$ is the Glaisher-Kinkelin constant. Zero-temperature counterterms, given in appendix \ref{appendix: zero temperature counterterms}, are used in the evaluation to cancel the temperature independent divergences. In order to simplify the formulae, we have also introduced notation following \cite{Kajantie:1995dw}:
\begin{equation}
L_b(\Lambda) \equiv 2 \log \left( \frac{e^{\gamma_E} \Lambda}{4 \pi T} \right)
\end{equation}

\begin{equation}
L_f(\Lambda) \equiv 2 \log \left( \frac{e^{\gamma_E} \Lambda}{\pi T} \right)
\end{equation}

\begin{equation}
c \equiv - \log \left(\frac{ 3 e^{\gamma_{E} / 2} A^6}{4 \pi}\right) = -0.348723...
\end{equation}

\begin{align}
\label{eq:gamma1_integrals}
  &\begin{aligned}
    \mathllap{\Gamma^{(1)}(\mathbf{0}) } & \approx \sigma + \delta \sigma + \frac{1}{2}g \mysumint_{Q} \frac{1}{Q^2} - \frac{1}{4} g \lambda \mysumint_{QR} \frac{1}{Q^2 R^4} - \frac{1}{6} g \lambda \mysumint_{QR} \frac{1}{Q^2 R^2 (Q+R)^2} \notag
  \end{aligned}\\
  &\begin{aligned}
    & \quad + \frac{1}{2} \delta g \mysumint_{Q} \frac{1}{Q^2} - \frac{1}{2} g(m^2 + \delta m^2)\mysumint_{Q} \frac{1}{Q^4} - \frac{1}{2} g \delta Z \mysumint_{Q} \frac{1}{Q^2} \notag
  \end{aligned}\\
  &\begin{aligned}
    & \quad + i y \mysumint_{\{ Q \}} \frac{\text{Tr}[\slashed{Q}]}{Q^2} + i \delta y \mysumint_{\{ Q \}} \frac{\text{Tr}[\slashed{Q}]}{Q^2} - y(m_\psi + \delta m_\psi) \mysumint_{\{ Q \}} \frac{\text{Tr}[\slashed{Q}\slashed{Q}]}{Q^4} \notag
  \end{aligned}\\
  &\begin{aligned}
    & \quad - i y \delta Z_f \mysumint_{\{ Q \}} \frac{\text{Tr}[\slashed{Q}\slashed{Q}\slashed{Q}]}{Q^4} - \delta y m_\psi \mysumint_{\{ Q \}} \frac{\text{Tr}[\slashed{Q}\slashed{Q}]}{Q^4}
  \end{aligned}\\
  &\begin{aligned}
    & \quad + y m_\psi^3 \mysumint_{\{ Q \}} \frac{\text{Tr}[\slashed{Q}\slashed{Q}\slashed{Q}\slashed{Q}]}{Q^8} + 2 y m_\psi \delta Z_f \mysumint_{\{ Q \}} \frac{\text{Tr}[\slashed{Q}\slashed{Q}\slashed{Q}\slashed{Q}]}{Q^6} \notag
  \end{aligned} \\
  &\begin{aligned}
    & \quad -i y^3 \mysumint_{\{ QR \}} \frac{\text{Tr}[\slashed{Q}\slashed{R}\slashed{Q}]}{Q^4(Q-R)^2R^2} + 2 y^3 m_\psi \mysumint_{\{ QR \}} \frac{\text{Tr}[\slashed{Q}\slashed{R}\slashed{Q}\slashed{Q}]}{Q^6(Q-R)^2R^2} \notag
  \end{aligned} \\
  &\begin{aligned}
    & \quad + y^3 m_\psi \mysumint_{\{ QR \}} \frac{\text{Tr}[\slashed{Q}\slashed{Q}\slashed{R}\slashed{R}]}{Q^4(Q-R)^2R^4} + \frac{1}{2} g y^2 \mysumint_{\{ QR \}} \frac{\text{Tr}[\slashed{Q}\slashed{R}]}{Q^2(Q-R)^4R^2} \notag
  \end{aligned} \notag \\ \notag \\ 
  &\begin{aligned} \label{eq:gamma1_integrated}
     &\approx \sigma + \frac{g T^2}{24} + \frac{1}{(4\pi)^2}\bigg[\frac{1}{6} g\lambda T^2\Big(\frac{1}{4 \epsilon} + \frac{1}{8} L_b(\Lambda) + 6\log(A) - \frac{1}{2} \gamma_E\Big)\\
      &\quad -\frac{1}{2} g m^2 L_b(\Lambda)\bigg] + \frac{m_\psi y T^2}{6} + \frac{1}{(4 \pi)^2}\bigg[4 y m_\psi^3 L_f(\Lambda) \\
      &\quad + \frac{g T^2 y^2}{24}(L_f(\Lambda) - 3L_b(\Lambda)) + \frac{m_\psi T^2 y^3}{6}(L_f(\Lambda) + 2L_b(\Lambda)) \bigg]
  \end{aligned}
\end{align}

\begin{align}
\label{eq:gamma2_integrals}
  &\begin{aligned}
    \mathllap{\Gamma^{(2)}(\mathbf{p,-p})} &\approx p^2 + m^2 + \delta m^2 + p^2 \delta Z + \frac{\lambda}{2} \mysumint_{Q} \frac{1}{Q^2} - \frac{g^2}{2} \mysumint_{Q} \frac{1}{Q^4} \left(1 - \frac{p^2}{Q^2} + \frac{4(\mathbf{p \cdot q})^2}{Q^4}\right) \notag
  \end{aligned}\\
  &\begin{aligned}
    & \quad - \frac{\lambda^2}{4} \mysumint_{QR} \frac{1}{Q^4 R^2} - \frac{\lambda^2}{6} \mysumint_{QR} \frac{1}{Q^2 R^2 (Q+R)^2} + \frac{\delta \lambda}{2} \mysumint_{Q} \frac{1}{Q^2}  \notag
  \end{aligned}\\
  &\begin{aligned}
    & \quad -g \delta g \mysumint_{Q} \frac{1}{Q^4} - \frac{\lambda}{2}(m^2 + \delta m^2) \mysumint_{Q} \frac{1}{Q^4} - \frac{\lambda}{2} \delta Z \mysumint_{Q} \frac{Q^2}{Q^4} \notag
  \end{aligned}\\
  &\begin{aligned}
    & \quad + g^2(m^2 + \delta m^2) \mysumint \frac{1}{Q^6} - 4y^2 \mysumint_{\{ Q \}} \frac{1}{Q^2} \left(1 - \frac{p^2}{Q^2} + \frac{2(\mathbf{p \cdot q})^2}{Q^4}\right) 
  \end{aligned}\\
  &\begin{aligned}
    & \quad - 2 i y^2 (m_\psi + \delta m_\psi) \mysumint_{\{ Q \}} \frac{\text{Tr}[\slashed{Q}\slashed{Q}\slashed{Q}]}{Q^6} + 2 y^2 \delta Z_f \mysumint_{\{ Q \}} \frac{\text{Tr}[\slashed{Q}\slashed{Q}\slashed{Q}\slashed{Q}]}{Q^6} \notag
  \end{aligned} \\
  &\begin{aligned}
    & \quad - 2 y \delta y \mysumint_{\{ Q \}} \frac{\text{Tr}[\slashed{Q}\slashed{Q}]}{Q^4} + y^2 m_\psi^2 \mysumint_{\{ Q \}} \frac{\text{Tr}[\slashed{Q} \slashed{Q} \slashed{Q}\slashed{Q}]}{Q^8} \notag
  \end{aligned} \\ 
  &\begin{aligned}
    & \quad + 2 y^2 m_\psi^2 \mysumint_{\{ Q \}} \frac{\text{Tr}[\slashed{Q}\slashed{Q}\slashed{Q}\slashed{Q}]}{Q^8} + y^4 \mysumint_{\{ QR \}} \frac{\text{Tr}[\slashed{Q}\slashed{Q}\slashed{R}\slashed{R}]}{Q^4(Q-R)^2R^4} \notag
  \end{aligned}\\
  &\begin{aligned}
    & \quad + 2 i g y^3 \mysumint_{\{ QR \}} \frac{\text{Tr}[\slashed{Q}\slashed{Q}\slashed{R}]}{Q^4(Q-R)^4R^2} + 2 y^4 \mysumint_{\{ QR \}} \frac{\text{Tr}[\slashed{Q}\slashed{Q}\slashed{Q}\slashed{R}]}{Q^6(Q-R)^2R^2} \notag
  \end{aligned}\\
  &\begin{aligned}
    &  \quad + \frac{1}{2} \lambda y^2 \mysumint_{\{ QR \}} \frac{\text{Tr}[\slashed{Q}\slashed{R}]}{Q^2(Q-R)^4R^2} \notag
  \end{aligned}\\ \notag \\ 
  &\begin{aligned}
     &\approx p^2 + m^2 + \frac{\lambda T^2}{24} - \frac{g^2 L_b(\Lambda)}{2(4 \pi)^2}\\
      &\quad + \frac{1}{(4\pi)^2}\bigg[\frac{\lambda^2 T^2}{12}\Big(\frac{1}{2\epsilon} + \frac{L_b(\Lambda)}{4}  - \gamma_E + 12 \log(A)\Big) -\frac{\lambda m^2}{2}  L_b(\Lambda)\bigg]\\
      &\quad + \frac{y^2}{6(4\pi)^2}\bigg[8p^2 + 16 T^2 \pi^2 + L_f(\Lambda) \Big(72 m_\psi^2 + 12 p^2 + y^2 T^2 + \frac{\lambda T^2}{4}\Big)\\
      &\quad + L_b(\Lambda) \Big(2 y^2 T^2 - \frac{3 \lambda T^2}{4}\Big)\bigg]
  \end{aligned}
\end{align}

\begin{align}
\begin{split}\label{eq:gamma3_integrals}
\Gamma^{(3)}(\mathbf{p,q,-p,-q}) \approx{} & g + \delta g - \frac{3}{2} g \lambda \mysumint_{Q} \frac{1}{Q^4} -2 i y^3 \mysumint_{Q} \frac{\text{Tr}[\slashed{Q}\slashed{Q}\slashed{Q}]}{Q^6} \\
& + 6 y^3 m_\psi \mysumint_{Q} \frac{\text{Tr}[\slashed{Q}\slashed{Q}\slashed{Q}\slashed{Q}]}{Q^8} 
\end{split} \\ \notag \\ 
\begin{split}\label{eq:gamma3_integrated}
\approx{} & g - \frac{3}{2(4 \pi)^2} g \lambda L_b(\Lambda) + \frac{24 y^3 m_\psi}{(4 \pi)^2} L_f(\Lambda)
\end{split}
\end{align}

\begin{align}
\begin{split}\label{eq:gamma4_integrals}
\Gamma^{(4)}(\mathbf{p,q,r,-p,-q,-r}) \approx{} & \lambda + \delta \lambda - \frac{3}{2} \lambda^2 \mysumint_{Q} \frac{1}{Q^4} + 6 y^4 \mysumint_{Q} \frac{\text{Tr}[\slashed{Q}\slashed{Q}\slashed{Q}\slashed{Q}]}{Q^8} 
\end{split} \\ \notag \\ 
\begin{split}\label{eq:gamma4_integrated}
\approx{} & \lambda - \frac{3}{2(4 \pi)^2} \lambda^2 L_b(\Lambda) + \frac{24 y^4}{(4 \pi)^2} L_f(\Lambda)
\end{split}
\end{align}

In evaluating these equations we expanded assuming $p \sim y T$. Performing the same calculation in the effective theory, and using the vanishing of scaleless integrals in dimensional regularisation, we find trivially that,

\begin{align}
\begin{split}\label{eq:gamma1_3d}
\Gamma_3^{(1)}(\mathbf{0}) \approx{} &  \sigma_3 + \delta \sigma_3
\end{split}\\
\begin{split}\label{eq:gamma2_3d}
\Gamma_3^{(2)}(\mathbf{p,-p}) \approx{} & p^2 + m_3^2 + \delta m_3^2
\end{split}\\
\begin{split}\label{eq:gamma3_3d}
\Gamma_3^{(3)}(\mathbf{p,q,-p,-q}) \approx{} &  g_3 + \delta g_3
\end{split}\\
\begin{split}\label{eq:gamma4_3d}
\Gamma_3^{(4)}(\mathbf{p,q,r,-p,-q,-r}) \approx{} &  \lambda_3 + \delta \lambda_3 \ \text{.}
\end{split}
\end{align}

Following ref.~\cite{Kajantie:1995dw}, we match the leading momentum dependence of the correlation functions through a field normalisation, as well as equate the 1 to 4 point correlation functions at zero momentum between the full and effective theories. Using these relationships we arrive at the expressions~\eqref{eq:dr result 1} to \eqref{eq:dr result 5} for the 3d EFT parameters. The following counterterms are required in the 3d theory to cancel against the temperature dependent poles from the full theory,

\begin{align}
\begin{split}
\delta \sigma_3 ={} &  \frac{g_3 \lambda_3}{24(4 \pi)^2 \epsilon} \ ,
\end{split}\\
\begin{split} \label{eq:3d mass counterterm}
\delta m_3^2 ={} &  \frac{\lambda_3^2}{24(4 \pi)^2 \epsilon} \ .
\end{split}
\end{align}

\section{Loop integrals} \label{appendix: loop integrals}
 In $d=3-2\epsilon$ dimensions, the loop integral measure is defined as
\begin{equation}
\int_p \equiv \left( \frac{e^\gamma \Lambda^2}{4 \pi} \right)^\epsilon \int \frac{d^{3-2\epsilon} p}{\left( 2 \pi \right)^{3-2\epsilon}}  \ ,
\end{equation}
where $\Lambda$ is the \MSbar~renormalisation scale. In $d=4-2\epsilon$, the loop integration measure is defined analogously. Sum integrals over loop momenta are defined as,
\begin{equation}
\mysumint_{P \text{ or } \{ P \}} \equiv T \sum_{n=-\infty}^{\infty} \int_p \ ,
\end{equation}
where thermal four momenta are denoted by uppercase letters, $P = (p_0, p)$, with spatial momenta $p$ and Matsubara frequencies $p_0 = 2 \pi T n$ for bosonic sum integrals (denoted by $P$) and $p_0 = (2n+1) \pi T $ for fermionic sum integrals (denoted by $\{P\}$).

Massless 1-loop bosonic integrals can be evaluated using the following master sum-integral,
\begin{align}
\mysumint_{P} \frac{(p_0^2)^{\beta}(p^2)^{\gamma}}{(P^2)^{\alpha}} = &  \left( \frac{e^{\gamma} \Lambda^2}{4 \pi} \right)^{\epsilon} \frac{2 T (2 \pi T)^{d - 2 \alpha + 2\beta + 2\gamma}}{(4 \pi)^{d/2}} 
 \nonumber \\
 &\qquad \cdot \frac{\Gamma(\frac{d}{2} + \gamma) \Gamma(-\frac{d}{2} + \alpha - \gamma)}{\Gamma(\frac{d}{2}) \Gamma(\alpha)} \zeta(-d+2\alpha-2\beta-2\gamma),
\end{align}
where $\zeta$ is the Riemann zeta function. Massless 1-loop fermion sum integrals can be expressed in terms of bosonic sum integrals using the relationship:
\begin{equation}
\mysumint_{\{Q\}}  =  2 \times \mysumint_{Q, T/2} - \mysumint_{Q}  \ ,
\end{equation}
where the first sum integral on the right side of the equation should have all temperatures substituted $T \rightarrow T/2$.

Massless 2-loop integrals can be evaluated using integration by parts techniques, reducing them to products of 1-loop sum integrals \cite{Ghisoiu:2012yk}. In changes of variables, the sum or difference of two fermionic variables is a bosonic variable. The following results were used:
\begin{align}
\begin{split}
\mysumint_{PQ} \frac{1}{P^2 Q^2 (P+Q)^2} = {} & 0
\end{split}\\
\begin{split}
\mysumint_{PQ} \frac{1}{P^4 Q^2 (P+Q)^2} = {} &  - \frac{1}{(d-2)(d-5)} \mysumint_{PQ} \frac{1}{P^4 Q^4}
\end{split}\\
\begin{split}
\mysumint_{PQ} \frac{1}{P^6 Q^2 (P+Q)^2} = {} & - \frac{4}{(d-2)(d-7)} \mysumint_{PQ} \frac{1}{P^4 Q^6}
\end{split}\\
\begin{split}
\mysumint_{PQ} \frac{1}{P^4 Q^4 (P+Q)^2} = {} & 0
\end{split}\\
\begin{split}
\mysumint_{\{PQ\}} \frac{1}{P^2 Q^2 (P-Q)^2} = {} & 0 .
\end{split}
\end{align}

\section{Zero temperature renormalisation} \label{appendix: zero temperature counterterms}

We have used dimensional regularisation, and adopt the following convention for the counterterm Lagrangian,
\begin{align}
    \mathscr{L}_\text{ct} &=  \frac{1}{2}\delta Z \partial_\mu \phi \partial^\mu \phi- \delta \sigma \phi - \frac{1}{2}\delta m^2\phi^2 - \frac{1}{3!}\delta g \phi^3 - \frac{1}{4!}\delta \lambda \phi^4 \\
    &\quad + \bar{\psi}(i\delta Z_f \slashed{\partial} - \delta m_\psi) \psi - \delta y \phi \bar{\psi} \psi \ ,
\end{align}
where $\phi$ and $\psi$ are renormalised fields. The total Lagrangian \eqref{eq:lagrangian} is then modified as $\mathscr{L} \to \mathscr{L} + \mathscr{L}_\text{ct}$. The 1-loop counterterms in $d=4-2\epsilon$ dimensions read
\begin{align}
\delta \sigma &= \frac{g m^2 - 8 y m_\psi^3}{2 (4 \pi)^2 \epsilon} ,
&
\delta m^2 &= \frac{g^2 + \lambda m^2 - 24 y^2 m_\psi^2}{2 (4 \pi)^2 \epsilon} ,
\\
\delta g &= \frac{3 g \lambda - 48 y^3 m_\psi}{2 (4 \pi)^2 \epsilon} ,
&
\delta \lambda &= \frac{3 \lambda^2 - 48 y^4}{2 (4 \pi)^2 \epsilon} ,
\\
\delta m_\psi &= \frac{y^2 m_\psi}{(4 \pi)^2 \epsilon} ,
&
\delta y &= \frac{y^3}{(4 \pi)^2 \epsilon} ,
\\
\delta Z &= - \frac{2 y^2}{(4 \pi)^2 \epsilon} ,
&
\delta Z_f &= - \frac{y^2}{2 (4 \pi)^2 \epsilon} .
\end{align}
The corresponding beta functions and anomalous dimensions are
\begin{align}
\beta_\sigma &= \frac{g m^2 - 8 y m_\psi^3 + 2 \sigma y^2}{(4 \pi)^2 } ,
&
\beta_{m^2} &= \frac{g^2 + \lambda m^2 + 4 y^2 m^2 - 24 y^2 m_\psi^2}{(4 \pi)^2 } ,
\\
\beta_g &= \frac{3 g \lambda - 48 y^3 m_\psi + 6 g y^2}{(4 \pi)^2 } ,
&
\beta_\lambda &= \frac{3 \lambda^2 - 48 y^4 + 8 \lambda y^2}{(4 \pi)^2 } ,
\\
\beta_{m_\psi} &= \frac{3 y^2 m_\psi}{(4 \pi)^2 } ,
&
\beta_y &= \frac{5 y^3}{(4 \pi)^2 } ,
\\
\gamma_\phi &= - \frac{2y^2}{(4\pi)^2} ,
&
\gamma_\psi &= - \frac{y^2}{2(4\pi)^2} .
\end{align}
The beta functions for the dimensionless couplings are corroborated by refs.~\cite{Cheng:1973nv, Baek:2012uj}.

Barred couplings explicitly demonstrating renormalisation scale invariance for the 3d EFT parameters in eqs.~\eqref{eq:dr result 1} to \eqref{eq:dr result 5} are defined as:
\begin{align}
\bar{\chi} ={} & \chi - \frac{1}{2} \beta_{\chi}^\text{b} L_b(\Lambda) - \frac{1}{2} \beta_{\chi}^\text{f} L_f(\Lambda),
\end{align}
where for each parameter, the full beta function is equal to the sum of the bosonic and fermionic parts $\beta = \beta^\text{b} + \beta^\text{f}$, with:

\begin{align}
\beta_{\sigma}^\text{b} &= \frac{g m^2}{(4 \pi)^2} ,
&
\beta_{\sigma}^\text{f} &= -\frac{8 y m_\psi^3}{(4 \pi)^2} + \frac{2 y^2 \sigma}{(4 \pi)^2},
\\
 \beta_{m^2}^\text{b} &= \frac{g^2}{(4 \pi)^2} + \frac{\lambda m^2}{(4 \pi)^2},
&
 \beta_{m^2}^\text{f} &=   -\frac{24 y^2 m_\psi^2}{(4 \pi)^2} + \frac{4 y^2 m^2}{(4 \pi)^2} ,
\\
\beta_{g}^\text{b} &= \frac{3 g \lambda}{(4 \pi)^2},
&
 \beta_{g}^\text{f} &=  -\frac{48 y^3 m_\psi}{(4 \pi)^2} + \frac{6 y^2 g}{(4 \pi)^2}  ,
\\
\beta_{\lambda}^\text{b} &= \frac{3 \lambda^2}{(4 \pi)^2}  ,
&
 \beta_{\lambda}^\text{f}  &=  -\frac{48 y^4}{(4 \pi)^2} + \frac{8 y^2 \lambda}{(4 \pi)^2}  ,
\\
\beta_{y}^\text{b} &=  - \frac{2 y^3}{(4 \pi)^2},
&
\beta_{y}^\text{f}  &=  \frac{7 y^3}{(4 \pi)^2} ,
\\
\beta_{m_\psi}^\text{b} &= -\frac{2 y^2 m_\psi }{(4 \pi)^2}  ,
&
\beta_{m_\psi}^\text{f}   &=  \frac{5 y^2 m_\psi }{(4 \pi)^2} .
\end{align}
For $m_\psi$ and $y$ there is some arbitrariness in this split, which has been chosen in order to minimise the presence of $\log(2)$ terms in the dimensional reduction relations, equations \eqref{eq:dr result 1} to \eqref{eq:dr result 5}.

\bibliographystyle{JHEP}
\bibliography{refs}

\end{document}